\documentclass[pdflatex,sn-mathphys]{sn-jnl}

\usepackage{subcaption}
\usepackage{array,amsmath,amssymb}
\usepackage{mathrsfs,tabularx}
\usepackage{mathtools}
\usepackage{gensymb,wasysym,marvosym}
\usepackage{natbib}
\usepackage{listings,booktabs,latexsym}

\newcommand{\drv}{\textrm{d}}
\newcommand{\defeq}{\vcentcolon=}

\jyear{2022}%

\theoremstyle{thmstyleone}%
\newtheorem{theorem}{Theorem}
\newtheorem{proposition}[theorem]{Proposition}%

\theoremstyle{thmstyletwo}%
\newtheorem{remark}{Remark}%

\theoremstyle{thmstylethree}%
\newtheorem{definition}{Definition}%

\begin{document}

\title[Polynomial Stochastic Dynamical Indicators]{Polynomial Stochastic Dynamical Indicators}


\author*[1]{\fnm{Massimiliano} \sur{Vasile}}\email{massimiliano.vasile@strath.ac.uk}
\author[1]{\fnm{Matteo} \sur{Manzi}}
\affil[1]{\orgdiv{Department of Mechanical \& Aerospace Engineering}, \orgname{University of Strathclyde}, \orgaddress{\street{75 Montrose Street}, \city{Glasgow}, \postcode{G1 1XJ}, \state{Scotland}, \country{UK}}}


\abstract{This paper introduces three types of dynamical indicators that capture the effect of uncertainty on the time evolution of dynamical systems. Two indicators are derived from the definition of Finite Time Lyapunov Exponents while a third indicator directly exploits the property of the polynomial expansion of the dynamics with respect to the uncertain quantities. The paper presents the derivation of the indicators and a number of numerical experiments that illustrates the use of these indicators to depict a cartography of the phase space under parametric uncertainty and to identify robust initial conditions and regions of practical stability in the restricted three-body problem.}



\maketitle

\section{Introduction}\label{sec:Introduction}
\textcolor{blue}{In \cite{szeb_uq} Victor Szebehely} underlines how dynamical models are approximation of real-world phenomena, and how initial conditions and parameters can be known with a finite degree of accuracy. The approximation in the modelling of natural phenomena and the degree of accuracy in model parameters and initial conditions are all aspects of the uncertainty in dynamical systems. A complete understanding of the evolution of a dynamical system requires a quantification of the effects of this uncertainty. More specifically, the goal is to compute a measure of the uncertainty in a given quantity of interest. In dynamical systems, the quantity of interest is often a function of the state variables at a given time and the value of the state variables is a function of the uncertain quantities in the dynamical model.

In the past two decades there has been a growing interest in developing methods for uncertainty quantification in dynamical systems. Broadly speaking, methods differ for the assumptions on the nature of the uncertainty, aleatory or epistemic, the way uncertainty is propagated and the quantity of interest is computed. 
A complete review of methods for uncertainty quantification in dynamical systems is out of the scope of this paper. Here we will focus on a rather large and popular class of these methods that uses polynomial expansions to model the dependency of the state variables or, directly, the quantity of interest, on the uncertain quantities. Among them it is worth mentioning methods that propagate high-order Taylor polynomials  \cite{massari2017, Palau2015}, Polynomial Chaos Expansions (PCE) \cite{bhusal2019,Ozen2017LongTP,GERRITSMA20108333,Schick} and Chebyshev polynomials \cite{Vasile:2019}. 

Often the study of dynamical systems makes use of indicators to identify chaotic behaviours,  diffusion phenomena and invariant and coherent structures (e.g., \cite{froe, Skokos_2009, DARRIBA_2012, Lega2016}). Among these indicators, the Finite-Time Lyapunov Exponent (FTLE) \citep{shadden2005} was recently proposed as an attempt to generalise the concept of Invariant Manifolds for non-autonomous dynamical systems \citep{haller2015}, and identify structures that separate qualitatively different dynamical regimes. Some applications can be found in \cite{gaw2007,SHORT2014592,hh2015,manzitopputo2021}. Other chaos indicators are the frequency map analysis \citep{Laskar1993}, the Mean Exponential Growth factor of Nearby Orbits (MEGNO); the Smaller Alignment Index (SALI); the Fast Lyapunov Indicator (FLI); the Dynamical Spectra of stretching numbers and the corresponding Spectral Distance and the Relative Lyapunov Indicator (RLI). A review of some of them can be found in \cite{rev_ind}. Another class of indicators are used to study dynamical systems driven by stochastic processes, from time series, e.g., \cite{Steeb2005,procaccia1983,TARNOPOLSKI2018834}.
However, to the best of our knowledge, there is no indicator that is designed to quantify the effect of uncertainty in the system dynamics. Commonly used chaos indicators, for example, would need to be re-computed for each realisation of the uncertain quantities and a statistics on their sensitivity to the variation of the uncertain quantities would need to be computed a posteriori from a Monte Carlo simulation. In this respect it is worth mentioning the work on the computation of Lyapunov Exponents of stochastic driven processes in \cite{Schomerus2002} and \cite{froyland2000}.

In this paper we propose three novel dynamical indicators that exploit the properties of polynomial expansions for uncertainty quantification. Two indicators generalise the concept of Finite Time Lyapunov Exponents to the case where the parameters of the dynamic model are uncertain. The third indicator directly relates the coefficients of the polynomial expansion to the rate at which an ensemble of trajectories, given by different realisations of the uncertain parameters, diffuses. All three indicators allow one to directly study the effect of uncertainty without the need to run a Monte Carlo simulation and recompute multiple times the value of the chaos indicators. Unlike previous works that aimed at differentiating deterministic chaos from the effect of stochastic processes \citep{rosso2007,poon2001,Panichi19} or identify particular types of motion from time series \citep{cincotta1999}, in this paper we propose indicators that quantify the effect of parametric uncertainty in the dynamic model. Furthermore, the third indicator, called pseudo-diffusion exponent in the following, is shown to be \textcolor{blue}{more} computationally advantageous as it does not require the derivation and propagation of the variational equations.

Three examples of known dynamical systems are used to illustrate the applicability of the three types of indicators to the construction of a cartography of the dynamics and the identification of regions, in the phase space, that are more or less sensitive to model uncertainty.
It will be shown that the new indicators provide results that are consistent with the FTLE, when the uncertainty is only in the initial conditions. When the uncertainty is in the parameters of the dynamic model, the new indicators allow one to identify behaviours that manifest only due to the presence of a parametric uncertainty. At the same time the new indicators, consistent with other chaos indicators in the literature, allow one to identify regions of regular and chaotic motion. However, unlike existing chaos indicators, the ones proposed in this paper provide additional information on these regions, including variance, skewness, and higher statistical moments, of the ensemble of trajectories induced by multiple realisations of the uncertain quantities. 

In particular we will show how the pseudo-diffusion exponent can be used to identify trajectories that are nearly insensitive to parametric uncertainty in the dynamics and others that, for the same initial conditions, manifest radically different behaviours for different realisations of the uncertain quantities.  

The paper is structured as follows. After introducing the problem that this paper is addressing and a brief summary of the background material, the paper introduces the definition and derivation of the three indicators. Then, the indicators are applied to three known dynamical systems where a model parameter is affected by uncertainty. A discussion section with computational cost and significance of the three indicators follows. Finally a section on the practical applicability of the indicators concludes the paper.

\section{Problem Statement}
\label{sec:problem}
In this work we consider a general dynamical system in the form:
\begin{equation}
\label{eq:diffeq}
    \frac{d \textbf{z}}{d t} = \mathbf{g}(t, \mathbf{p}, \mathbf{z})
\end{equation}
with initial conditions:
\begin{equation}
\label{eq:icond}
    \mathbf{z}(t=t_0)=\mathbf{z}_0
\end{equation}
where $t$ is the time, $\textbf{z}:[t_0,t_f]\rightarrow \mathbb{R}^{n}$ is the state of the system and $\textbf{p}\in\Omega\subset \mathbb{R}^{n_p}$ is a vector of uncertain model parameters. In the general case, both $\mathbf{p}$ and $\mathbf{z}_0$ are uncertain quantities and similar in nature. The vector function $\mathbf{g} : [t_0,t_f]\times\mathbb{R}^{n_p}\times\mathbb{R}^{n}\longrightarrow \mathbb{R}^{n}$ is the dynamic model.

The objective is to derive a scalar quantity $\alpha(\mathbf{z},t):\mathbb{R}^{n}\times[t_0,t_f]\rightarrow \mathbb{R}$ that measures the divergence of the trajectories of system \eqref{eq:diffeq}, belonging to an ensemble $\Phi(t,\mathbf{p})=\{\mathbf{z}(t,\mathbf{p})\lvert\forall \mathbf{p}\in\Omega \wedge t\in[t_0,t_f]\}$ induced by multiple realisations of $\mathbf{p}$. We want also to quantify the uncertainty in the distance between a realisation $\mathbf{z}$ and the mean value of all the realisations in the ensemble at a given time $t_f$. We can quantify this uncertainty by computing the integral:
\begin{equation}\label{eq:exp_in}
    \mathbb{E}(\delta(t_f)<\epsilon)=\int_{\Omega} I(\|\mathbf{z}(t_f)-\hat{\mathbf{z}}(t_f)\|<\epsilon)w(\mathbf{p})d\mathbf{p}
\end{equation}
where $\delta=\|\mathbf{z}(t_f)-\hat{\mathbf{z}}(t_f)\|$, $I$ is the indicator function, $\epsilon$ is a threshold value and $\hat{\mathbf{z}}(t_f)$ is the mean value of the state variables at time $t_f$, or:
\begin{equation}
\label{eq:mean_z}
    \hat{\mathbf{z}}(t_f)=\frac{\int_{\Omega} \mathbf{z}(t_f)w(\mathbf{p})d\mathbf{p}}{\int_{\Omega} w(\mathbf{p})d\mathbf{p}}
\end{equation}
The function $w$ can represent the distribution of $\mathbf{p}$ over $\Omega$. In this case \eqref{eq:exp_in} is a probability and \eqref{eq:mean_z} an expected value. 


\section{Background Material}\label{sec:meth}
In this section we recall some basic material that is required to derive the dynamical indicators proposed in this paper. In particular we will focus on polynomial expansions to propagate the uncertainty in $\mathbf{p}$ through system \eqref{eq:diffeq}. Thus we will first briefly introduce both intrusive and non-intrusive Polynomial Chaos Expansions.

Two of the indicators are derived from Finite Time Lyapnov Exponents, hence, a subsection will introduce the concept of FTLE. Finally, one dynamic indicator is based on the idea of anomalous diffusion in stochastic systems, therefore, the last subsection will present some basic concepts of anomalous diffusion.

\subsection{Polynomial Expansions}
A popular technique to study the dependency of a dynamical system on a set of uncertain quantities is  Polynomial Chaos Expansions. The idea is to represent the state vector $\mathbf{z}$ as a truncated expansion in the orthogonal polynomials $\Psi_i(\mathbf{p})$ of the uncertain quantities $\mathbf{p}$:
\begin{equation}
\label{eq:pce101}
\mathbf{z}(t, \mathbf{p}) \approx \sum_{i=0}^m \mathbf{c_i}(t) \Psi_i(\mathbf{p})
\end{equation}
where $\mathbf{c}_i(t)$ are time dependent coefficients. The $\Psi_i$ terms define a set of orthogonal polynomials up to degree $m$ \citep{orthpoli}. The orthogonality condition is formalised as follows:
\begin{equation}
\label{eq:orthogonality}
    \langle \Psi_j, \Psi_k \rangle = \int_{\Omega} \Psi_j(\mathbf{p}) \Psi_k (\mathbf{p}) w(\mathbf{p}) \drv \mathbf{p} = \mathbb{E}[\Psi_j, \Psi_k] \neq 0 \Leftrightarrow j = k
\end{equation}
where $\langle \cdot, \cdot \rangle$ is a shorthand of the inner product. As mentioned before when the $w$ is a distribution  \eqref{eq:orthogonality} defines the expectation operator associated to $w$. Because of the polynomial nature of the terms appearing in \eqref{eq:orthogonality}, it is straightforward to compute the non-zero terms. Then, given a particular weight function $w(\mathbf{p})$, one can use the following three terms recursion relation given in \cite{ttr} to create stabilised univariate orthogonal polynomials:
\begin{equation}
\label{eq:ttr}
    \Psi_{i+1}(p) = \Psi_{i}(p)(p-A_i) - \Psi_{i-1}(p)B_i, \ \ \ \ A_i = \frac{\mathbb{E}[p  \Psi^2_{i}]}{\mathbb{E}[\Psi^2_{i}]}, \ \ \ \ B_i = \frac{\mathbb{E}[\Psi^2_{i}]}{\mathbb{E}[\Psi^2_{i-1}]} 
\end{equation}

In the case in which more than one source of uncertainty is present, it is still possible to construct orthogonal multivariate polynomials via tensor product rules \citep{FEINBERG201546}. Note that while the method proposed in this paper is applicable to any orthogonal polynomial constructed with \eqref{eq:ttr} in all the examples in this paper Chebyshev basis functions of the second kind are used together with the associated weight function $w(\mathbf{p})$.

By substituting the approximation given by \eqref{eq:pce101} in  \eqref{eq:diffeq}, one gets:
\begin{equation}
\label{eq:pce2}
\frac{d \textbf{z}}{\drv t} = \frac{d}{d t} \sum\limits_{i=0}^m \mathbf{c_i}(t) \Psi_i(\mathbf{p}) = \sum\limits_{i=0}^m \mathbf{\dot{c}_i}(t) \Psi_i(\mathbf{p}) = \mathbf{g}(t, \mathbf{p}, \mathbf{z}) 
\end{equation}
and by making use of the intrusive Galerkin method, one obtains the following:
\begin{equation}
\label{eq:pce3}
\left\langle \sum\limits_{i=0}^m \mathbf{\dot{c}_i}(t) \Psi_i(\mathbf{p}), \Psi_k(\mathbf{p}) \right\rangle = \left\langle \mathbf{g}(t, \mathbf{p}, \mathbf{z}), \Psi_k(\mathbf{p}) \right\rangle
\end{equation}
from which the time variation of the coefficients can be derived:
\begin{equation}
\label{eq:finalpce}
\dot{\mathbf{c}}_k(t) = \frac{\left\langle \mathbf{g}(t, \mathbf{p}, \mathbf{z}), \Psi_k(\mathbf{p}) \right\rangle}{\left\langle \Psi_k(\mathbf{p}), \Psi_k(\mathbf{p}) \right\rangle} 
\end{equation}

The integrals at numerator of the right hand side of \eqref{eq:finalpce} need to be computed numerically, in the general case, while the integrals at denominator can be pre-computed analytically. Gauss quadrature rules \citep{FEINBERG201546} can be used to approximate the integrals at numerator, as follows:
\begin{equation}
\label{eq:quadrature}
\begin{array}{l}
\left\langle \mathbf{g}(t, \mathbf{p}, \mathbf{z}), \Psi_k(\mathbf{p}) \right\rangle = \int_{\Omega} \mathbf{g}(t, \mathbf{p}, \mathbf{z}(\mathbf{p})) \Psi_k (\mathbf{p}) w(\mathbf{p}) d \mathbf{p} \approx \\
\approx\sum\limits_{j_1=1}^N...\sum\limits_{j_i=1}^N...\sum\limits_{j_n=1}^N W_{j_1}...W_{j_i}...W_{j_n} \mathbf{g}(t, \mathbf{p}_{j_i}, \mathbf{z}(\mathbf{p}_{j_i})) \Psi_k (\mathbf{p}_{j_i})
\end{array}
\end{equation}
where $W_{j_i}$ and $\mathbf{p}_{j_i}$ are respectively the $N$ quadrature weights and abscissa points along each dimension $i$. 
Sparse quadrature schemes \cite{smolyak} can be used to reduce the computational complexity of the numerical integrals with the increase in the number of dimensions.

The initial value of the coefficients $\mathbf{c}_k(t=0)$ is found by projecting the initial conditions $\mathbf{z}_0$:
\begin{equation}
    \label{eq:pceic}
    \mathbf{c}_k(t=0)  = \frac{\left\langle \mathbf{z_0}, \Psi_k(\mathbf{p}) \right\rangle}{\langle \Psi_k(\mathbf{p}), \Psi_k(\mathbf{p}) \rangle}
\end{equation}
which greatly simplifies in the case in which the initial state is deterministic (i.e., none of the components of $\mathbf{z_0}$ are components of $\mathbf{p}$): the only non-zero coefficient is $\mathbf{c_0}$, the one associated to the degree-zero polynomial of the orthogonal basis, whose value is the one of the deterministic initial condition.

Up to this point PCEs are simply a way to represent the state of the system $\mathbf{z}$ with a polynomial expansion of the parameters $\mathbf{p}$ and propagate this expansion forward in time.
Thus regardless of whether $\mathbf{p}$ is an uncertain quantity with an associated probability distribution $w$ or a simple parameter defined on a parameter space $\Omega$, \eqref{eq:cg_pce} provides a way to propagate the polynomial forward in time. 

Furthermore, \eqref{eq:pceic} can be applied at any time $t$ to calculate a polynomial expansion of the state variables with respect to the uncertainty variables. In this case \eqref{eq:pceic} reads:
\begin{equation}
    \label{eq:pceni}
    \hat{\mathbf{c}}_k(t)  = \frac{\left\langle \mathbf{z}(t,\mathbf{p}), \Psi_k(\mathbf{p}) \right\rangle}{\langle \Psi_k(\mathbf{p}), \Psi_k(\mathbf{p}) \rangle}
\end{equation}

In both \eqref{eq:pceic} and \eqref{eq:pceni} the integral at denominator can be computed analytically, one time before the calculation of the coefficients. The integral at numerator of \eqref{eq:pceni} can \textcolor{blue}{be} solved numerically as in \eqref{eq:quadrature}:
\begin{equation}
\label{eq:quadrature_z}
\begin{array}{l}
\left\langle \mathbf{z}(t, \mathbf{p}) \Psi_k(\mathbf{p}) \right\rangle = \int_{\Omega} \mathbf{z}(t, \mathbf{p}) \Psi_k (\mathbf{p}) w(\mathbf{p}) d \mathbf{p} \approx \\
\approx\sum\limits_{j_1=1}^N...\sum\limits_{j_i=1}^N...\sum\limits_{j_n=1}^N W_{j_1}...W_{j_i}...W_{j_n} \mathbf{z}(t, \mathbf{p}_{j_i}) \Psi_k (\mathbf{p}_{j_i})
\end{array}
\end{equation}
The polynomial expansion computed with \eqref{eq:pceni} is called non-intrusive because one needs only samples of the state vector $\mathbf{z}(t,\mathbf{p})$ at time $t$ for different realisations of $\mathbf{p}$. These samples can be obtained from the direct forward integration of the equations of motion.

The use of a non-intrusive computation of the coefficients of the polynomial expansion is advantageous when the dynamical model is not directly accessible, the state vector is available through observations or, as it will be explained in section \ref{sec:results}, if the integration of system \eqref{eq:pce_stm} becomes problematic due to the presence of singularities or discontinuities in the uncertainty space.
In this case restart mechanisms like the ones proposed in \cite{greco:2020,manziaas2020} and \cite{ozen} can be effectively used to improve the propagation of the polynomial expansion.
In this paper, however, we will not consider these restart mechanisms and we will show the use of \eqref{eq:pceni} instead of \eqref{eq:finalpce} to compute two of the indicators.

Since the interest is to exploit the evolution of the coefficients of a polynomial expansion and not to exactly propagate a particular probability distribution, the weight $w$ and basis functions $\Psi$ can be arbitrarily chosen to make the numerical integration of \eqref{eq:quadrature} efficient. In the following we will consider the components of $\mathbf{p}$ to be independent and $\Omega$ to be a orthotope. Furthermore, integral \eqref{eq:quadrature} is performed after the change of coordinates:
\begin{equation}
    p_i=\frac{(b_i-a_i)}{2}\xi_i+\frac{b_i+a_i}{2}\;\;\; i=1,...,n
\end{equation}
with $p_i\in[a_i,b_i]$ and $\mathbf{\xi}\in[-1,1]^n$ so that:
\begin{equation}
\label{eq:quadrature_norm}
\int_{\Omega} \mathbf{g}(t, \mathbf{p}, \mathbf{z}(\mathbf{p})) \Psi_k (\mathbf{p}) w(\mathbf{p}) d \mathbf{p}= \frac{\prod_i^n(b_i-a_i)}{2^n}\int_{[-1,1]^n} \mathbf{g}(t, \mathbf{\xi}, \mathbf{z}(\mathbf{\xi})) \Psi_k (\mathbf{\xi}) w(\mathbf{\xi}) d \mathbf{\xi}
\end{equation}

In this section we derived the expansion, intrusive or non-intrusive, of $\mathbf{z}$ in orthogonal polynomials of $\mathbf{p}$. Other forms of uncertainty quantification in the literature, like Taylor series expansions, for example, do not use orthogonal polynomials. However, in the definition of the stochastic dynamical indicators we will exploit the orthogonality of the polynomials. Thus, while, in principle, any polynomial representation of $\mathbf{z}$ is applicable, before computing the stochastic indicators one would need to transform the polynomial expansion into orthogonal basis as suggested in \cite{Iosto2022}. 

Note also that the use of Taylor expansions to derive dynamical indicators was already proposed in \cite{Palau2015}. However, the approach introduced in this paper differs from the one in \citep{Palau2015} in two important ways: i) in this paper we use the evolution of the coefficients of the polynomials to directly define the indicators and ii) the indicators proposed in this paper quantify the effect of uncertainty in the parameters defining the dynamic model. This later point is of particular importance because, as it will be explained in the remainder of the paper, the primary utility of the indicators proposed in this work is to study the effect of the uncertainty in the dynamic model.  

\subsection{Finite-Time Lyapunov Exponent}
Following \cite{milani} Section 2.3 we now briefly recall the definition of Finite-Time Lyapunov Exponents.
We start from the definition of \textcolor{blue}{the} variational equations in the deterministic settings:
\begin{equation}
\label{eq:pce_stm}
d \textbf{z} (t, \textbf{p}) \approx \frac{\partial \textbf{z}(t, \textbf{p})}{\partial \mathbf{z_0}} \drv \mathbf{z_0}
\end{equation}
The FTLE emerges from the spectral analysis of the Cauchy--Green (CG) Strain Tensor:
\begin{equation}\label{eq:Delta}
    \Delta = \Phi^T \Phi
\end{equation}
where $\Phi$ is the state transition matrix of the system. From it, the definition of Finite-Time Lyapunov Exponent \citep{shadden2005} is given by:
\begin{equation}\label{eq:ftle}
\sigma(\mathbf{z}(t_f,\mathbf{p}))=\frac{1}{t_f-t_0} \log{\sqrt{\lambda_{max}(\mathbf{z}(t_f,\mathbf{p}))}}
\end{equation}
where $t_f$ is the time interval associated to the propagation, starting at $t_0$, and $\lambda_{max}$ is the maximum eigenvalue of the Cauchy--Green Strain Tensor.

\subsection{Random Walks, Mean Square Displacement and Diffusion}
A random-walk is a stochastic process that defines a path made of random steps. Steps can have random direction, random length and be taken at random times. One of the best known random-walks is Brownian motion. Brownian motion can be well described by a Weiner process $W_t$ with independent steps and each step taken from a normal distribution $\mathcal{N}(0,t)$ with zero mean and variance $Dt$.
\label{sec:random_walk}
\begin{equation}
    x(t)-x_0=\sqrt{2D}W_t
\end{equation}
\begin{equation}\label{eq:Weiner1D}
    \large\langle(x(t)-x_0)^2\large\rangle=W_t^2=2Dt
\end{equation}
where $D$ is the diffusion coefficient. In normal diffusion the exponent of the time $t$ is one, however, some stochastic processes can diffuse faster or slower (e.g. fractionated Brownian motion or Levy processes) \citep{ALVES2016392}. Thus in the general case one can write:
\begin{equation}\label{eq:anomal_diff}
    \large\langle(x(t)-x_0)^2\large\rangle\approx Kt^{\alpha}
\end{equation}
where $K$ is a constant and $\alpha$ is the diffusion exponent. In the next section we will make use of \eqref{eq:anomal_diff} to derive an indicator that relates the coefficients of the polynomial expansion to the diffusion exponent.

\section{Stochastic Dynamical Indicators}
\label{sec:SDI}
In this section we introduce and define three different types of stochastic dynamical indicators, or SDIs. The first one is a simple quantification of the uncertainty in the FTLE induced by multiple realisation of the uncertain parameter vector $\mathbf{p}$. The second type of indicator is an extension of the idea of FTLE that measures the divergence of two polynomial expansions of neighbouring trajectories. The third type measures the degree of diffusion of an ensemble of trajectories induced by multiple realisation of the uncertain quantities.

\subsection{Stochastic Finite-Time Lyapunov Exponents}
In this section we will develop two types of Stochastic Finite-Time Lyapunov Exponents. 
The first type replaces the FLTE with the statistical moments quantifying the uncertainty in the FTLE. If the dynamics depends on some uncertain quantities, the Strain Tensor in \eqref{eq:Delta} is a random matrix with entries that are a function of the realisations of the uncertain quantities. Thus one could study the ensemble of matrices and derive a statistics over the realisations of the eigenvalues.
An approach to derive the statistical moments of the FTLE can be found in \cite{Schomerus2002}. In \cite{Schomerus2002} the authors considered the case of a one-dimensional dynamical system driven by a random potential and built the statistical moments of the FTLE by computing the moments of the components of the matrix $\partial \mathbf{z}(t_f)(t)/\partial\mathbf{z}_0$.  In what follows, instead, we will use a polynomial chaos expansion of the FTLE with respect to the uncertain vector $\mathbf{p}$. By sampling the uncertain space $\Omega$, one can directly construct the PCE expansion of the FTLE $\sigma$ defined in \eqref{eq:ftle}:
    \begin{equation}
    \label{eq:sigma_pce}
    \sigma(\mathbf{z}(t_f,\mathbf{p})) \approx \sum_{k=0}^m \sigma_k(t_f) \Psi_k(\mathbf{p})
\end{equation}
where the coefficients $\sigma_k(t_f)$ are computed by projection:
\begin{equation}
    \label{eq:projected_sigma}
    \sigma_k(t_f)  = \frac{\langle\sigma(\mathbf{z}(t_f,\mathbf{p})), \Psi_k(\mathbf{p})\rangle }{\langle \Psi_k(\mathbf{p}), \Psi_k(\mathbf{p}) \rangle}
\end{equation}
\begin{definition}\label{def:2}
We call Stochastic Finite-Time Lyapunov Exponents type 1 the statistical moments of the FTLE derived from expansion \eqref{eq:sigma_pce}:
\begin{equation}
    \alpha_1^1=\sigma_0
\end{equation}
\begin{equation}
\label{eq:variance_pce}
    \alpha_1^2=\sum_{k=1}^m\sigma_k^2\langle\Psi_k, \Psi_k \rangle
\end{equation}
For all higher moments one can use the multinomial expansion and pre-calculate the integrals of the basis functions:
\begin{equation}\alpha_1^m=\sum_{\lvert\mathbf{k}\rvert=m}\binom{m}{k_1,k_1,...,k_q} \left\langle\prod_{j=1}^q\Psi_j^{k_j} \right\rangle\prod_{j=1}^q \sigma_j^{k_j}
\end{equation}
where $\langle\prod_{j=1}^q\Psi_j^{k_j} \rangle$ can be pre-computed given a set of basis function and associated distribution function, and $\lvert\mathbf{k}\rvert=m$ means all the combination of indexes $k_j$ such that the sum is equal to $m$.
\end{definition}
\begin{remark}
From the definition of Stochastic Finite-Time Lyapunov Element type 1, it is clear that the same procedure described above can be applied to any other deterministic indicator to derive their statistical moments as a function of the distribution of $\mathbf{p}$. 
\end{remark}

For the second type, as in the deterministic settings, we start from the hypervolume $d\mathbf{z}^Td\mathbf{z}$ and compute the time evolution of its expectation $\mathbb{E}(d\mathbf{z}^Td\mathbf{z})$. 
\begin{proposition}
Given two solutions of system \eqref{eq:diffeq} and assuming that each solution can be expanded in the same orthogonal basis functions $\Psi(\mathbf{p})$ of the uncertain parameter vector $\mathbf{p}$, and given the distribution function $w(\mathbf{p})$, the expected value of the square difference of the two solutions can be approximated with:
\begin{equation}\label{eq:exp_dzdz}
    \mathbb{E}(d\mathbf{z}^Td\mathbf{z})\approx\sum_{i=0}^m d\mathbf{z}_0^T\left( \frac{\partial \mathbf{c_i}}{\partial \mathbf{z_0}}^T \frac{\partial \mathbf{c_i}}{\partial \mathbf{z_0}} \right)d\mathbf{z}_0 \langle\Psi_i, \Psi_i \rangle
\end{equation}
\end{proposition}
\begin{proof}
Given the two solutions $\mathbf{z}(\mathbf{p},t:\mathbf{z}_0)$ and $\mathbf{\hat{z}}(\mathbf{p},t:\mathbf{\hat{z}}_0)$, with initial conditions $\mathbf{z}_0$ and $\mathbf{\hat{z}}_0$, under the assumption that the solutions can be expanded in the same basis functions $\Psi_i$, we can write:
\begin{equation}
d \textbf{z}= \textbf{z}(t,\textbf{p}:\mathbf{z}_0)- \mathbf{\hat{z}}(t,\textbf{p}:\mathbf{\hat{z}}_0) \approx \sum_{i=0}^m \mathbf{c_i} \Psi_i - \sum_{i=0}^m \mathbf{\hat{c}_i} \Psi_i   
\end{equation}
and from \eqref{eq:pce_stm} calling $d\mathbf{z}_0=\mathbf{z}_0-\mathbf{\hat{z}}_0$ we have:
\begin{equation}
d \textbf{z} \approx\sum_{i=0}^m d \mathbf{c_i} \Psi_i \approx \sum_{i=0}^m \frac{\partial \mathbf{c_i}}{\partial \mathbf{z_0}} \drv \mathbf{z_0}\Psi_i 
\end{equation}
 from which, computing the expected value of the square of the final offset, we obtain:
 
\begin{equation}
\label{eq:expdzdz}
\begin{split}
\mathbb{E}(d \textbf{z}^Td \textbf{z})  \approx \int_{\Omega} \sum_{i=0}^m \sum_{j=0}^m d \mathbf{c_i} d \mathbf{c_j} \Psi_i \Psi_j  w(\mathbf{p}) \drv \mathbf{p}&=\\  &=  
\sum_{i=0}^m d \mathbf{c_i}^T d\mathbf{c_i} \langle \Psi_i, \Psi_i \rangle\approx \\ &\approx
\sum_{i=0}^m d\mathbf{z}_0^T\left( \frac{\partial \mathbf{c_i}}{\partial \mathbf{z_0}}^T \frac{\partial \mathbf{c_i}}{\partial \mathbf{z_0}} \right)d\mathbf{z}_0 \langle\Psi_i, \Psi_i \rangle
\end{split}
\end{equation}
\end{proof}

We now derive an equivalent definition of variational equations \eqref{eq:pce_stm} but in the coefficients of the PCE expansion of $d\mathbf{z}$.
\begin{proposition}
Given a dynamical system \eqref{eq:diffeq}, the following set of equations describes a Polynomial Chaos Expansion-based generalisation of the variational equations:
\begin{equation}\label{eq:var_c}
    \frac{\partial}{\partial t} \frac{\partial \mathbf{c_k}}{\partial \mathbf{z_0}} =\frac{1}{\langle \Psi_k, \Psi_k \rangle} \left\langle \frac{\partial \textbf{g}}{\partial \textbf{z}} \sum_{i=0}^m \left( \frac{\partial \mathbf{c_i}}{\partial \mathbf{z_0}} \Psi_i \right), \Psi_k  \right\rangle
\end{equation}
\end{proposition}

\begin{proof}
The following holds for ``smooth'' dynamics:
\begin{equation}
\label{eq:smooth}
\frac{\partial}{\partial t} \left[ \frac{\partial \textbf{z}}{\partial \mathbf{z_0}} (t, \textbf{p}, \mathbf{z_0}) \right] = \frac{\partial}{\partial \mathbf{z_0}} \left[ \frac{\partial \textbf{z}}{\partial t} (t, \textbf{p}, \mathbf{z_0}) \right]
\end{equation}
where the term in brackets is explicitly given by:
\begin{equation}
\label{eq:mooth2}
\frac{\partial \textbf{z}}{\partial t} (t, \textbf{p}, \mathbf{z_0}) = \textbf{g}(t, \textbf{p}, \mathbf{z}) = \textbf{g}(\textbf{z}(\mathbf{z_0},t,\textbf{p}), \textbf{p}, t)
\end{equation}
Therefore, we can write:
\begin{equation}
\label{eq:varMilani}
\frac{\partial}{\partial t} \left[ \frac{\partial \textbf{z}}{\partial \mathbf{z_0}} (t, \textbf{p}, \mathbf{z_0}) \right] = \frac{\partial}{\partial \mathbf{z_0}} \textbf{g}(\textbf{z}(\mathbf{z_0},t,\textbf{p}), \textbf{p}, t)
\end{equation}

By using the PCE decomposition, the second term in Eq. \eqref{eq:varMilani} leads to:
$$
\frac{\partial}{\partial \mathbf{z_0}} \textbf{g}(\textbf{z}(\mathbf{z_0},t,\textbf{p}), \textbf{p}, t) \approx \frac{\partial}{\partial \mathbf{z_0}} \textbf{g}(\textbf{z}(\mathbf{c_1}(t, \mathbf{z_0}), \dots, \mathbf{c_m}(t, \mathbf{z_0}), \textbf{p}, t), \textbf{p}, t) = 
$$
\begin{equation}
\label{eq:pcevar1}
= \frac{\partial \textbf{g}}{\partial \textbf{z}} \sum_{i=0}^m \left( \frac{\partial \textbf{z}}{\partial \mathbf{c_i}} \frac{\partial \mathbf{c_i}}{\partial \mathbf{z_0}} \right) = \frac{\partial \textbf{g}}{\partial \textbf{z}} \sum_{0=1}^m \left( \frac{\partial \mathbf{c_i}}{\partial \mathbf{z_0}} \Psi_i \right)
\end{equation}

while the first term of Eq. \eqref{eq:varMilani} leads to:
\begin{equation}
\label{eq:pcevar2}
\frac{\partial}{\partial t} \left[ \frac{\partial }{\partial \mathbf{z_0}} \sum_{i=0}^m \mathbf{c_i}(t, \mathbf{z_0}) \Psi_i(\textbf{p}) \right] = \frac{\partial}{\partial t} \sum_{i=0}^m \frac{\partial \mathbf{c_i}}{\partial \mathbf{z_0}} \Psi_i
\end{equation}

By putting Eqs. \eqref{eq:pcevar1} and \eqref{eq:pcevar2} back into Eq. \eqref{eq:varMilani} one gets:
\begin{equation}
\label{eq:varMilanipce}
\frac{\partial}{\partial t} \sum_{i=0}^m \frac{\partial \mathbf{c_i}}{\partial \mathbf{z_0}} \Psi_i = \frac{\partial \textbf{g}}{\partial \textbf{z}} \sum_{i=0}^m \left( \frac{\partial \mathbf{c_i}}{\partial \mathbf{z_0}} \Psi_i \right)
\end{equation}
and, by making use of the orthogonality condition, one arrives at the following result:
\begin{equation}
\label{eq:varManzi}
\frac{\partial}{\partial t} \frac{\partial \mathbf{c_k}}{\partial \mathbf{z_0}} =\frac{1}{\langle \Psi_k, \Psi_k \rangle} \left\langle \frac{\partial \textbf{g}}{\partial \textbf{z}} \sum_{i=0}^m \left( \frac{\partial \mathbf{c_i}}{\partial \mathbf{z_0}} \Psi_i \right), \Psi_k  \right\rangle
\end{equation}
\end{proof}

As discussed for the deterministic formulation in \cite{gaw2007}, in order to compute the variation of the coefficients $\mathbf{c}_i$, it is possible to propagate a regularly spaced grid of tracers with the same dimension as the phase space. In fact, the spectral harmonics of the generalised State Transition Matrix appearing in Eq. \eqref{eq:pce_stm} consist of partial derivatives which can be computed via central differencing of neighboring tracers, making use of the following second-order approximation:

\begin{equation}
    \label{eq:finitediff}
    \frac{\partial (c_{ki})_{t_0}^{t_f}(\mathbf{z})}{\partial z_j} \approx  \frac{(c_{ki})_{t_0}^{t_f}(\mathbf{z} + \Delta \mathbf{z}_j) - (c_{ki})_{t_0}^{t_f}(\mathbf{z} - \Delta \mathbf{z}_j)}{2 \Delta z_j}
\end{equation}
with $\Delta \mathbf{z}_j = [0, \dots, 0, \Delta z_j, 0, \dots, 0]$.
This methodology greatly reduces the computational cost associated to the generalisation of the variational equations, as it is for the deterministic case. While the accuracy of the computation of the CG tensor degrades with this approach, the authors in \cite{shadden2005} points out that: ``finite differencing may unveil Lagrange Coherent Structures more reliably than obtaining derivatives of the flow analytically''.

From \eqref{eq:exp_dzdz}, one can now introduce the Cauchy-Green Tensor $\Delta^c_{ii}$ of the coefficients $\mathbf{c}_i$:
\begin{equation}
\label{eq:cg_pce}
\Delta^c_{ii}:=\left( \frac{\partial \mathbf{c_i}}{\partial \mathbf{z_0}}^T \frac{\partial \mathbf{c_i}}{\partial \mathbf{z_0}} \right)
\end{equation}

\begin{definition}\label{def:1}
From the spectral decomposition of $\Delta^c_{ii}$ one can derive the maximum eigenvalue $\lambda_{ii,max}$ and then compute the corresponding exponent:
\begin{equation}\label{eq:sftle}
\alpha_2^i \defeq \frac{1}{t_f-t_0} \ln\sqrt{\lambda_{ii,max}}
\end{equation}
We call Stochastic Finite-Time Lyapunov Exponents type 2 the quantity $\alpha_2^i$ defined in \eqref{eq:sftle}.
The quantity $\alpha_2^i$ gives an indication of the deformation of the hypervolume $d\mathbf{c}_i^Td\mathbf{c}_i$. We can understand this deformation as the difference in the way two polynomial expansions of $\mathbf{z}$ with respect to $\mathbf{p}$, for two infinitesimally close initial conditions, evolve in time. 
\end{definition}

\begin{remark}
Note that $d\mathbf{c}_i^Td\mathbf{c}_i$ measures the hypervolume defined by each coefficient vector of the polynomial expansion. Thus the definition of $\alpha_2^i$ suggests the following:
\begin{itemize}
    \item if the polynomial expansion converges rapidly with $m$, high order coefficients will be small and so is expected to be the hypervolume $d\mathbf{c}_i^Td\mathbf{c}_i$
    \item if two trajectories, starting from infinitesimally close initial conditions evolve very differently in time, the polynomial expansion with respect to $\mathbf{p}$ is also expected to evolve very differently. This descends from the definition of the time derivative of the coefficients $\mathbf{c}_i$ that depends on $\mathbf{g}$ which is a function of $\mathbf{z}$.
    \item if multiple independent realisations of $\mathbf{p}$ induce trajectories that evolve very differently in time, a higher order expansion will be needed to properly represent $\mathbf{z}$ at a given time $t$, furthermore if two trajectories, starting from infinitesimally close initial conditions evolve very differently in time, one would expect a significant difference in the time evolution of high order coefficients $\mathbf{c}_i$.
\end{itemize}
We will expand further on these three points in the discussion section of the paper.
\end{remark}
Indicators in Definition \ref{def:2} will be called SFTLE1 in the remainder of this paper while indicators in Definition \ref{def:1} will be called SFTLE2.
Indicators SFTLE1 give the probability distribution of the FTLE in \eqref{eq:ftle} as a function of the distribution of the uncertain parameter vector $\mathbf{p}$. Indicators SFTLE2, instead, give a measure of the divergence of the coefficients of the polynomial model of the distribution of the solution $\mathbf{z}(t,\mathbf{p})$ as a function of a variation of the initial condition $\mathbf{z}_0$.
It should be noted how the eigenvectors associated to the parameter-dependent Cauchy--Green strain tensor are also characterised by a probability distribution. This implies that the direction of maximum strain is not deterministic, and there may be configurations in which there is an abrupt change of the maximum strain direction for different realisations of the uncertain parameter.








\subsection{Pseudo-Diffusion Exponent}
\label{sec:diff_exp}
In order to derive the third indicator we start from the idea, introduced in section \ref{sec:random_walk}, that in a generic random-walk process, the expected value of the square of the displacement is proportional to $Kt^{\alpha}$.
In the univariate case, by using Eq. \eqref{eq:variance_pce} and exploiting the orthogonality of the basis functions, one can write the expected value of the square displacement as:
\begin{equation}\label{eq:diff_expo_1}
    \kappa_2=\large\langle z-z_0,z-z_0 \large\rangle=\large\langle  \left(\sum_{i=0} c_i\psi_i-c_0\right)^2 \large\rangle=\sum_{i=1} s_i c_i^2
\end{equation}
with $s_i= \langle \psi_i,\psi_i\rangle$. One can now equate $\kappa_2$ to $Kt^{\alpha}$ to obtain:
\begin{equation}
\label{eq:diff_expo_2}
    \sum_{i=1} s_i c_i^2(t)=Kt^{\alpha}
\end{equation}
The left hand side is the variance of $z$ at time $t$, which, for $\alpha=1$, is consistent with the fact that for a one-dimensional Brownian motion the second statistical moment of the Mean Square Displacement (MSD) is $2Dt+z_0^2$, with $2D=K$ the diffusion coefficient, and the MSD is equal to the second cumulant of the Gaussian distribution characterising the Brownian motion.
This suggests that by looking at the variation of the coefficients of the polynomial, one can study the dynamical character of a system. Since the coefficients are subject to the same dynamic equations, see \eqref{eq:cg_pce}, they reflect the same evolution of the state.
The evolution of the coefficients can be derived in other ways, for example via an algebra on the space of the polynomials \cite{greco:2020, Palau2015}. As long as the state can be expressed as an expansion in orthogonal polynomials one can derive Eq. \eqref{eq:diff_expo_1}.
\begin{proposition}
The coefficient $\alpha$ in expression \eqref{eq:diff_expo_2} can be approximated by:
\begin{equation*}
  \alpha\approx\tilde{\alpha}=\frac{\log{\left (\sum_{i=1}^m s_i c_i^2(t) +1\right)}}{\log t}
\end{equation*}
\end{proposition}
\begin{proof}
Take the logarithm of both sides of expression \eqref{eq:diff_expo_2} after adding a 1:
\begin{equation}\label{eq:k2_alpha2}
  \log{\left (\sum_{i=1}^m s_i c_i^2(t) +1\right)}=b+\alpha\log t+\log\left(1+\frac{1}{Kt^{\alpha}}\right)
\end{equation}
with $b=\log K$, which can be re-written as:
\begin{equation}\label{eq:k2_alpha3}
  \frac{\log{\left (\sum_{i=1}^m s_ic_i^2(t) +1\right)}}{\log t}=\frac{b}{\log t}+\alpha+\frac{\log\left(1+\frac{1}{Kt^{\alpha}}\right)}{\log t}
\end{equation}
and for large $t$ can be approximated by:
\begin{equation}\label{eq:k2_alphatilde}
  \alpha\approx\tilde{\alpha}=\frac{\log{\left (\sum_{i=1}^m s_i c_i^2(t) +1\right)}}{\log t}
\end{equation}
\end{proof}
\begin{definition}
In the following we call the quantity $\tilde{\alpha}$ defined in Eq.\ \eqref{eq:k2_alphatilde}, \textit{pseudo-diffusion exponent}.
If $\mathbf{z}$ is a vector of dimension $n$ then one can write the covariance matrix: 
\begin{equation}\label{eq:cov_v}
\mathbf{C}_v =\left[\begin{array}{ccc}
\sum_{i=1}^m s_i c_{1,i}^2(t)&...&\sum_{i=1}^m s_i 
c_{1,i}(t)c_{n,i}(t)\\
...&...&...\\
...&\sum_{i=1}^m s_i c_{j,i}^2(t)&...\\
...&...&...\\
\sum_{i=1}^m s_i c_{n,i}(t)c_{1,i}(t)&...&\sum_{i=1}^m s_i c_{n,i}^2(t)\\
\end{array}
\right]
\end{equation}
In this case, given that the covariance matrix is positive semi-defined, the pseudo diffusion exponent can be computed as follows:
\begin{equation}\label{eq:k2_alpha4_orig}
  \tilde{\alpha}=\frac{\log{\left (\sum_{i=1} \lambda_i(\mathbf{c}(t)) +1\right)}}{\log t}
\end{equation}
where $\lambda_i$ is the $i^\text{th}$ eigenvalue of $\mathbf{C}_v$.
If only one component along the diagonal of the matrix $\mathbf{C}_v$ is considered for the computation of $\tilde{\alpha}$ we call the indicator $\tilde{\alpha}_j$ with the subscript corresponding to the $j^\text{th}$ component. In this case the indicator gives the rate of expansion of the projection of the polynomial along one axis only. In the remainder of the paper we will use the following slightly different definition:
\begin{equation}\label{eq:k2_alpha4}
  \tilde{\alpha}=\frac{\log{\left (\sqrt{\max_{i=1} \lambda_i(\mathbf{c}(t)) }+1\right)}}{\log t}
\end{equation}
\end{definition}
Note that both intrusive and non-intrusive propagation methods can be used to compute the coefficients of the polynomials at time $t$. However, in all the examples in this paper the pseudo-diffusion exponent will be computed with a non-intrusive computation of the coefficients.


\section{Numerical Experiments}
\label{sec:results}
In this section we test the applicability of all three types of indicators to the study of three well-known problems: the uncertain perturbed pendulum, the uncertain double gyre, the uncertain Circular Restricted Three-body Problem. For each of these problems we will construct a cartography and, by inspection, will analyse the characteristics of some notable trajectories. All simulations start at $t_0=0$. The code for all the simulations and analyses in this section was written in Matlab R2021b and was run on a laptop i7, 2.80GHz, in Windows 10 pro.  
In all the cases in this section, the expectation $\mathbb{E}$ defined in \eqref{eq:exp_in}) is computed by taking 100 uniformly distributed random samples of the uncertain vector $\mathbf{p}$ and computing the corresponding polynomial chaos model at time $t_f$. Numerical quadrature formulae were computed with 9 abscissa points and associated weights. From the experiments on the problems in this section a higher number of abscissa points did not bring any significant change in the indicators and we could reduce the abscissa points to 6 without important degradations of the results.

\subsection{The Uncertain Perturbed Pendulum}
The motion of a periodically-perturbed pendulum can be written as in \cite{Palau2015}:
\begin{equation}
\label{eq:pendulum}
\Ddot{x} = (a \cos 5 t -1 ) \sin x \\
\end{equation}
or as an equivalent system of first order differential equations:
\begin{equation}
    \label{eq:1orderode}
    \dot{\mathbf{z}} = \frac{\drv}{\drv t}
    \begin{array}{c}
\begin{bmatrix}
x \\
v_x
\end{bmatrix}
\end{array}
=
    \begin{array}{c}
\begin{bmatrix}
v_x \\
(a \cos 5 t -1 ) \sin x
\end{bmatrix}
\end{array}
=
\mathbf{g}(\mathbf{z}, p, t)
\end{equation}
with $p=a$ an uncertain parameter. One can then write the Jacobian of system \eqref{eq:1orderode}:
\begin{equation}
\label{eq:pendulumjacob}
\frac{\partial \textbf{g}}{\partial \textbf{z}}
=
\begin{array}{cc}
\begin{bmatrix}
0, & 1 \\
\cos x (a \cos 5 t - 1),  & 0
\end{bmatrix}
\end{array}
\end{equation} 
The uncertain parameter $a$ is defined over the interval $a\in[2.5-0.25,\; 2.5+0.25]$, with known or unknown distribution, dynamics \eqref{eq:pendulum} becomes uncertain and its evolution depends on the realisations of $a$. Thus we expanded the state variables in Chebyshev polynomials of parameter $a$, up to degree 4, and used the definition of the three indicators SFTLE1, SFTLE2 and $\tilde{\alpha}$ to study the evolution of the system.  

\begin{figure}[htb]
    \centering
     \begin{subfigure}[b]{0.48\textwidth}
         \centering
         \includegraphics[width=1.0\textwidth]{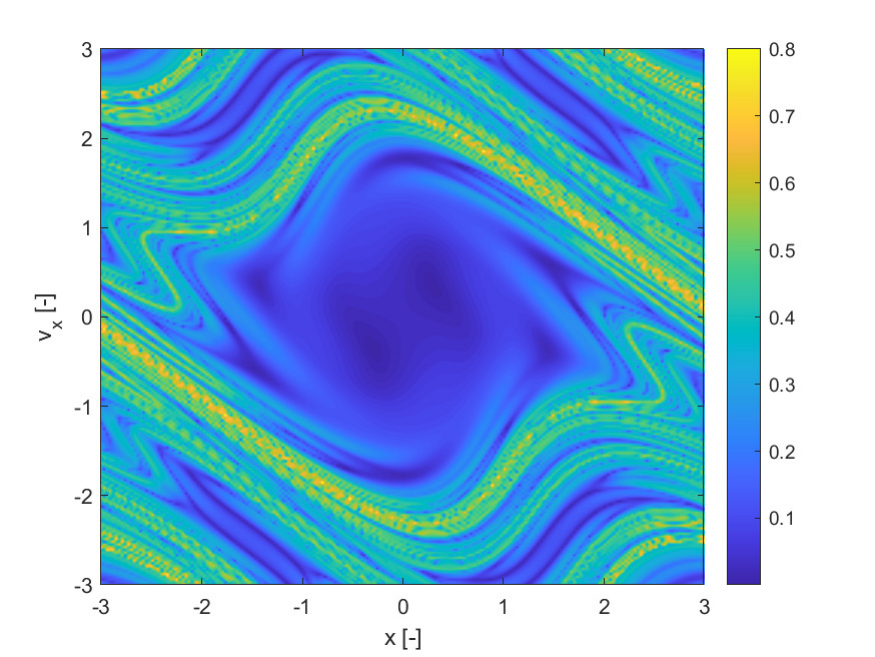}
         \caption{FTLE\label{fig:pendulum_sftlea}}
     \end{subfigure}
     \hfill
     \begin{subfigure}[b]{0.48\textwidth}
         \centering
         \includegraphics[width=1.0\textwidth]{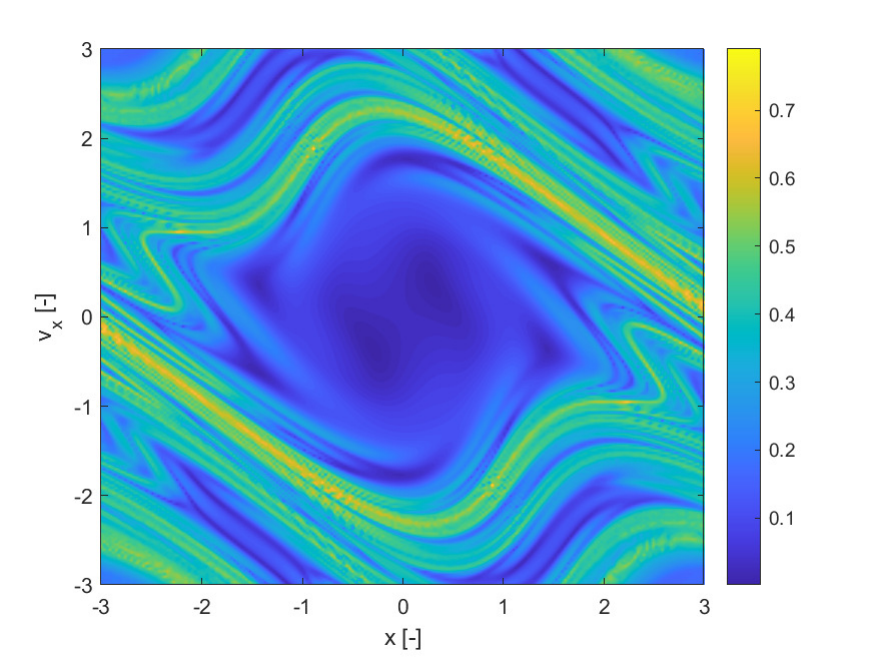}
         \caption{$\alpha_1^1$\label{fig:pendulum_sftleb}}
     \end{subfigure}
     \begin{subfigure}[b]{0.48\textwidth}
         \centering
         \includegraphics[width=1.0\textwidth]{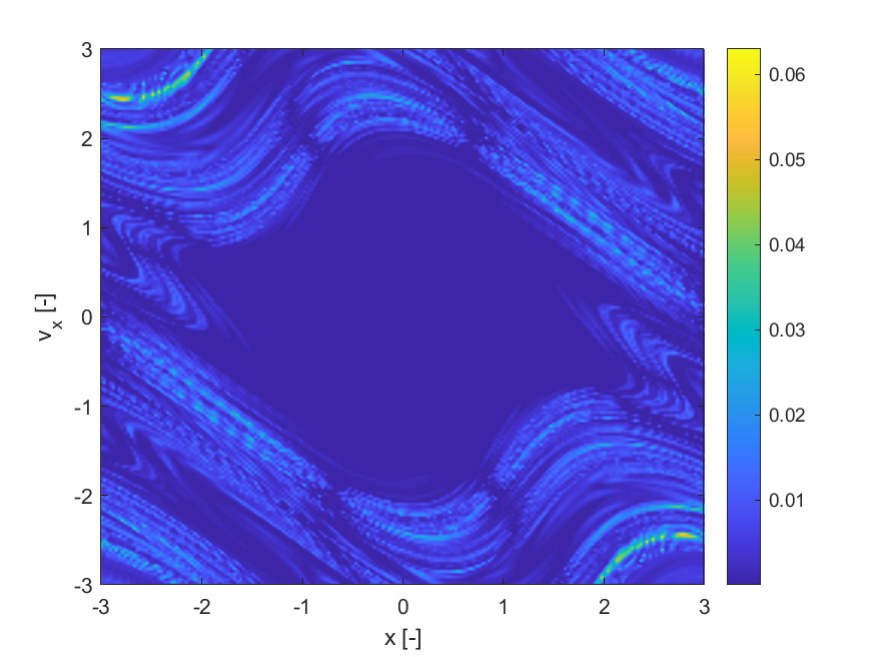}
         \caption{$\alpha_1^2$\label{fig:pendulum_sftlec}}
     \end{subfigure}
         \hfill
     \begin{subfigure}[b]{0.48\textwidth}
         \centering
         \includegraphics[width=1.0\textwidth]{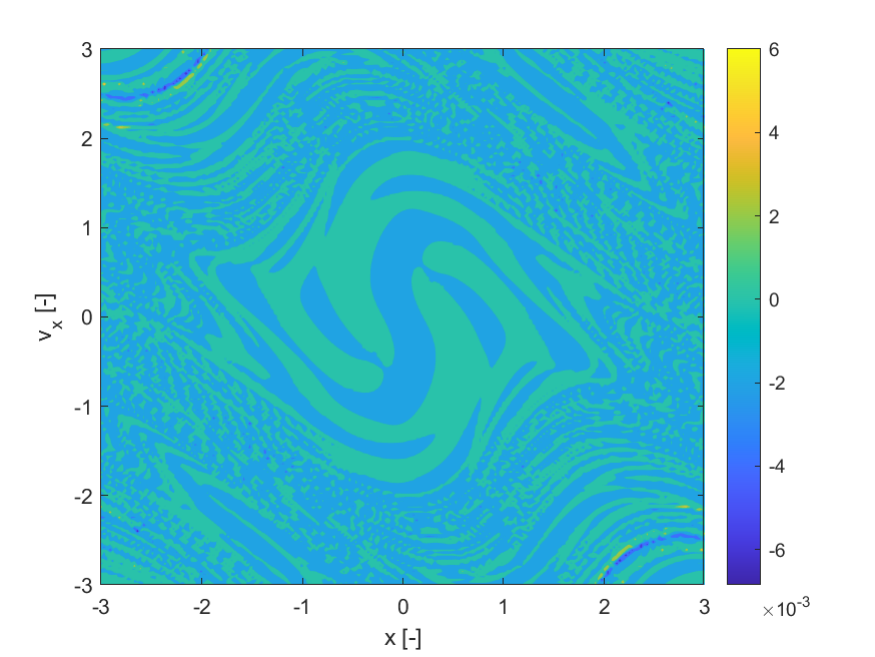}
         \caption{$\alpha_1^3$\label{fig:pendulum_sftled}}
     \end{subfigure}
        \caption{SFTLE type 1 scalar fields of the perturbed pendulum for $t_f= 10$.}
        \label{fig:pendulum_sftle}
\end{figure}

\begin{figure}[htb]
    \centering
     \begin{subfigure}[b]{0.48\textwidth}
         \centering
         \includegraphics[width=1.0\textwidth]{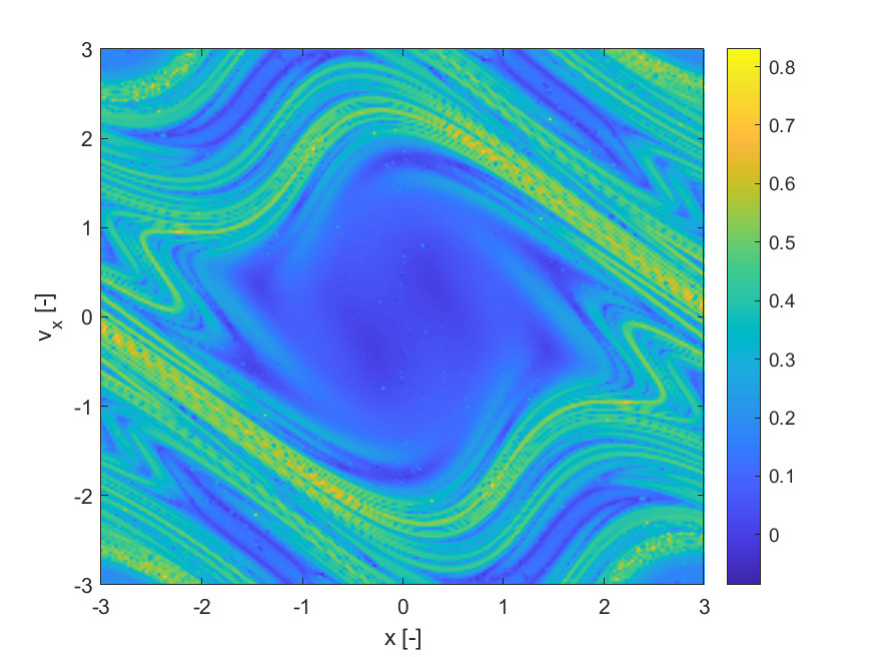}
         \caption{$\alpha_2^1$\label{fig:pendulum_sftle2a}}
     \end{subfigure}
     \begin{subfigure}[b]{0.48\textwidth}
         \centering
         \includegraphics[width=1.0\textwidth]{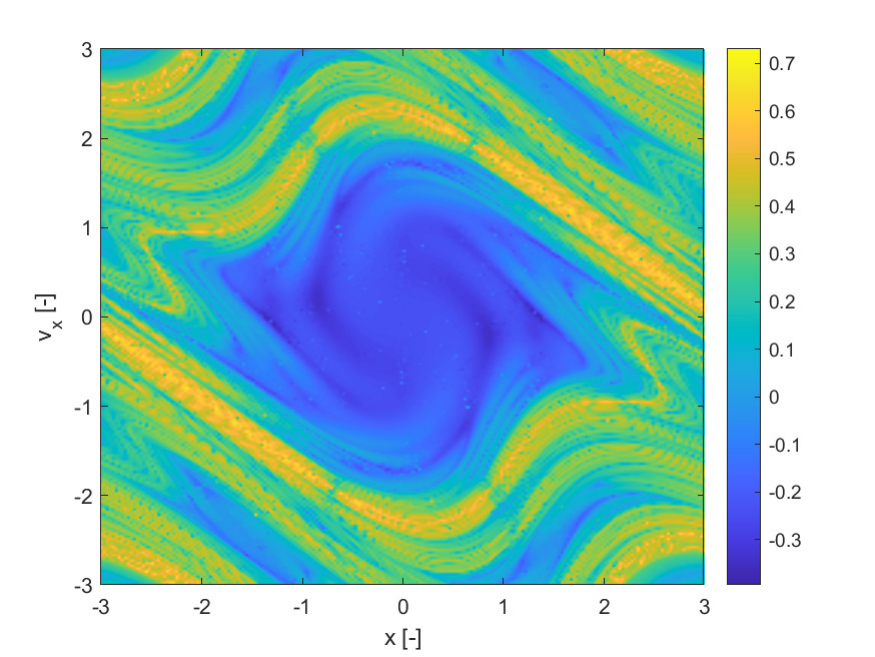}
         \caption{$\alpha_2^2$\label{fig:pendulum_sftle2b}}
     \end{subfigure}
         \hfill
     \begin{subfigure}[b]{0.48\textwidth}
         \centering
         \includegraphics[width=1.0\textwidth]{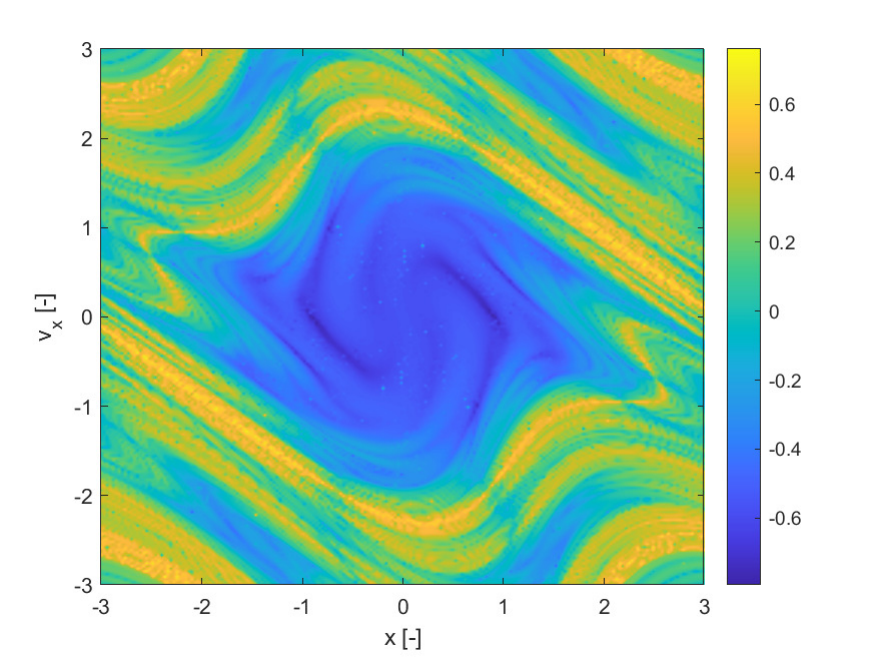}
         \caption{$\alpha_2^3$\label{fig:pendulum_sftle2c}}
     \end{subfigure}
        \caption{SFTLE type 2 scalar fields of the perturbed pendulum for $t_f = 10$.}
        \label{fig:pendulum_sftle2}
\end{figure}

All differential equations, were propagated forward in time for $t_f = 10$, with an explicit adaptive Runge-Kutta method of order 4/5 with absolute tolerance and relative tolerance respectively $10^{-10}$ and $10^{-9}$. 
The three indicators were computed over a uniform grid of $200\times200$ initial conditions over the domain $x \in [-3, 3]$, $v_x \in [-3, 3]$. The finite increment for he calculation of both the FTLE and SFTLE is $\Delta z_j = 1 \cdot 10^{-7}$.

Figure \ref{fig:pendulum_sftlea} shows the deterministic FTLE for $a=2.5$ while Figure \ref{fig:pendulum_sftleb} shows $\alpha_1^1$ for $a$ uncertain. Although the magnitude of the two indicators is slightly different, they present the same structures, as to be expected given that $\alpha_1^1$ is an average value over the realisations of $a$. Figure \ref{fig:pendulum_sftlec} represents the variance of the FTLE due to the uncertainty in $a$ and Figure \ref{fig:pendulum_sftled} the skewness.
Because the $\sin()$ is an odd function, the mapping $(x, v_x) \mapsto (-x, -v_x)$ is a symmetry of \eqref{eq:1orderode} and, because of this, the results given in Figure \ref{fig:pendulum_sftle} are characterised by a central symmetry with respect to the origin. Note however that Figure \ref{fig:pendulum_sftled} clearly shows that 
the realisations of the state vector at time $t_f$ are positively or negatively skewed depending on the initial conditions. Thus SFTLE1 provides different information on the distribution of the FTLE depending on the order of the indicator.

Figures \ref{fig:pendulum_sftle2} show the SFTLE2 from order 1 to 3. In this case all three indicators show the same structures but with very different ranges. To be noted that as the order increases the regions where the indicators are negative become more negative. This implies that the higher the coefficient $c$ the more two expansions starting from neighbouring initial conditions tend to behave similarly.

 \begin{figure}[htb]
     \centering
     \subfloat[$\tilde{\alpha}$]{{\includegraphics[width=0.5\textwidth]{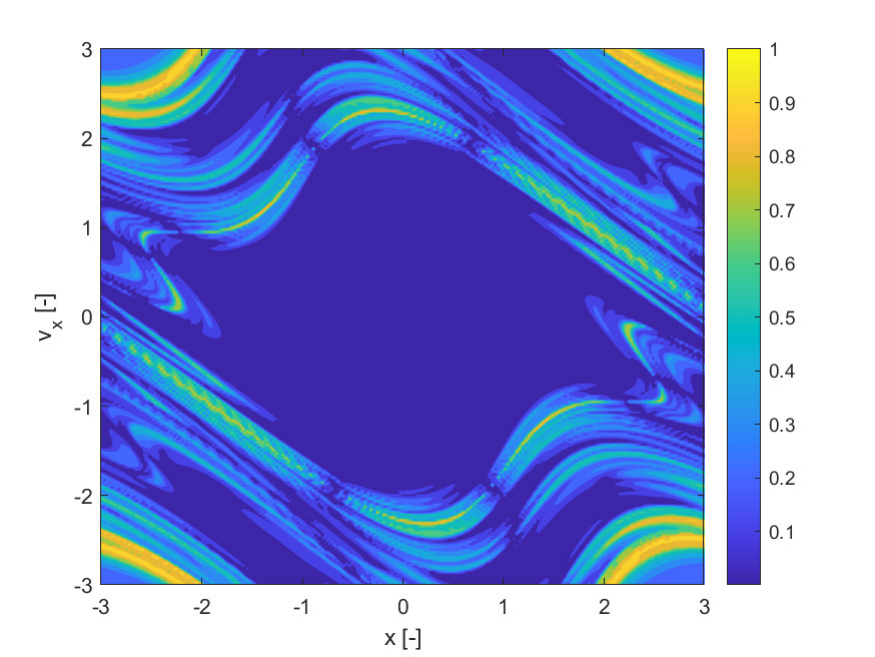}}}
     \subfloat[$\log_{10}{\tilde{\alpha}}$]{{\includegraphics[width=0.5\textwidth]{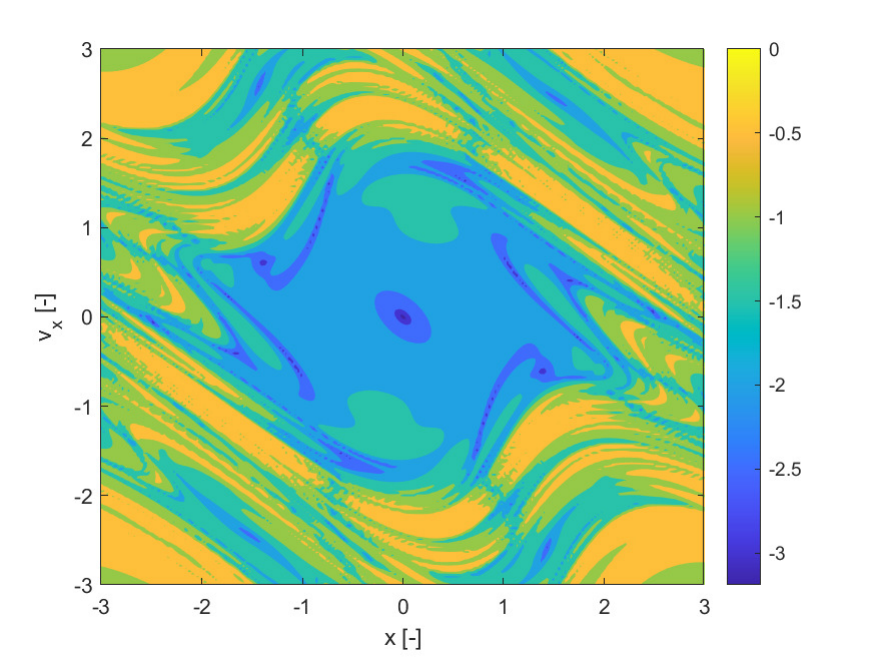}}}\\
     \subfloat[$\mathbb{E}_{0.1}$]{{\includegraphics[width=0.5\textwidth]{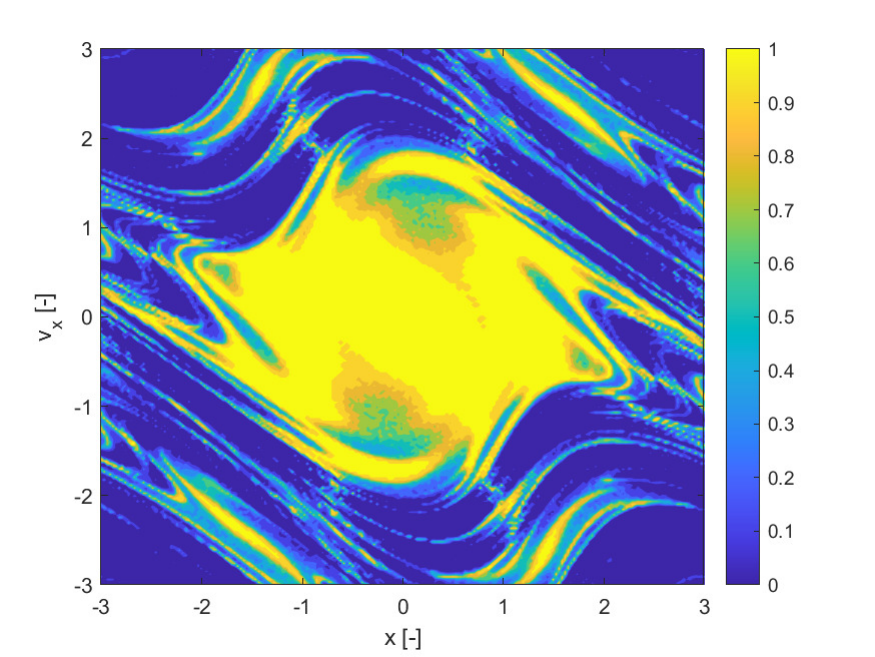}}}
     \subfloat[Skewness of $x$]{{\includegraphics[width=0.5\textwidth]{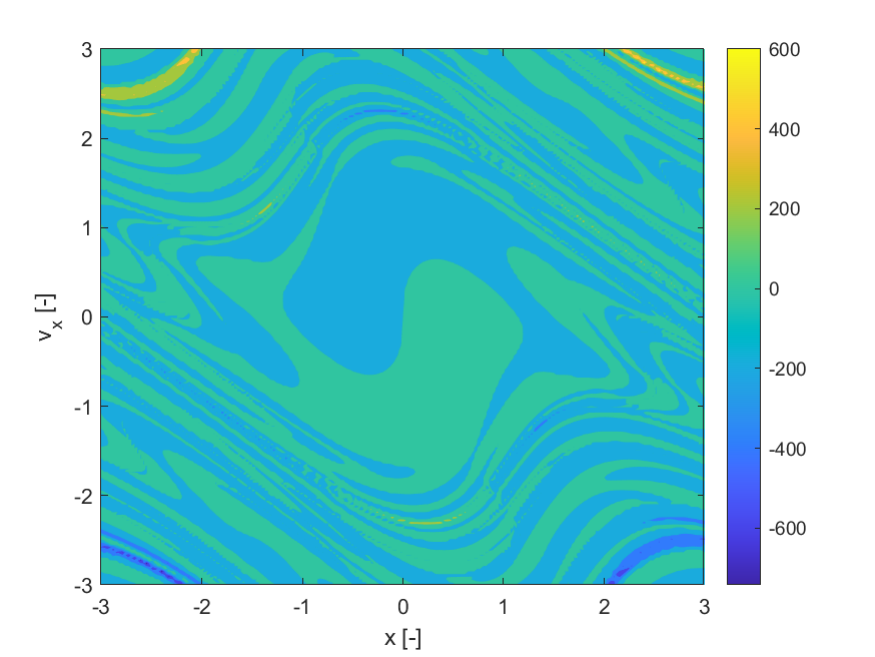}}}\\     
     \caption{Pseudo diffusion exponent field for the uncertain perturbed pendulum model.}
     \label{fig:pendulum_sdi}
\end{figure}

Figures \ref{fig:pendulum_sdi} show the pseudo diffusion exponent field together with the probability of the trajectories in the ensemble to remain within a distance $\epsilon=0.1$ from the mean at time $t_f$ (see Eq.\ \eqref{eq:exp_in}) and the skewness of the ensemble of trajectories induced by multiple realisations of the uncertain parameter $a$. 
The skewness is computed only for the state component $x$. For multivariate problems one would need to compute the skewness vector \cite{KOLLO20082328} and then reduce it to a scalar indicator. This computation will be addressed in future work.
Figure \ref{fig:pendulum_sdi}b shows the $\log10$ of Figure \ref{fig:pendulum_sdi}a. Also this indicator identifies the same structures as SFTLE1 and 2 and the associated skewness is consistent with Figure \ref{fig:pendulum_sftle}d. Figure \ref{fig:pendulum_sdi}c provides some additional information. First it is interesting to note that it is the negative image of Figure \ref{fig:pendulum_sdi}a which is consistent with the idea that $\tilde{\alpha}$ provides a measure of the diffusion of the trajectories. Then \ref{fig:pendulum_sdi}c highlights how some sets of initial conditions, yellow regions, are weakly sensitive to the uncertainty in $a$.

 \begin{figure}[htb]
     \centering
     \subfloat[]{{\includegraphics[width=0.54\textwidth]{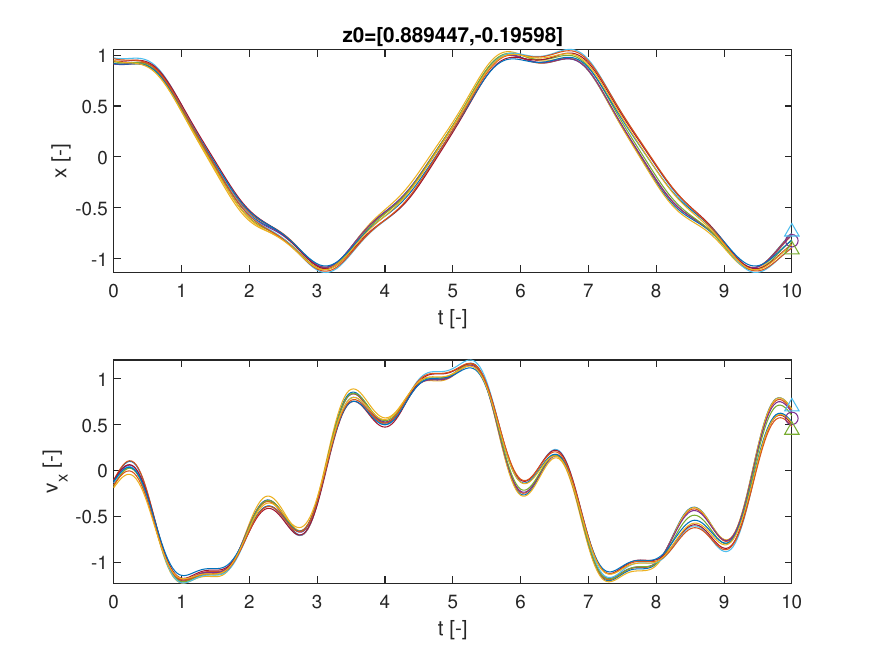}}}
     \subfloat[]{{\includegraphics[width=0.54\textwidth]{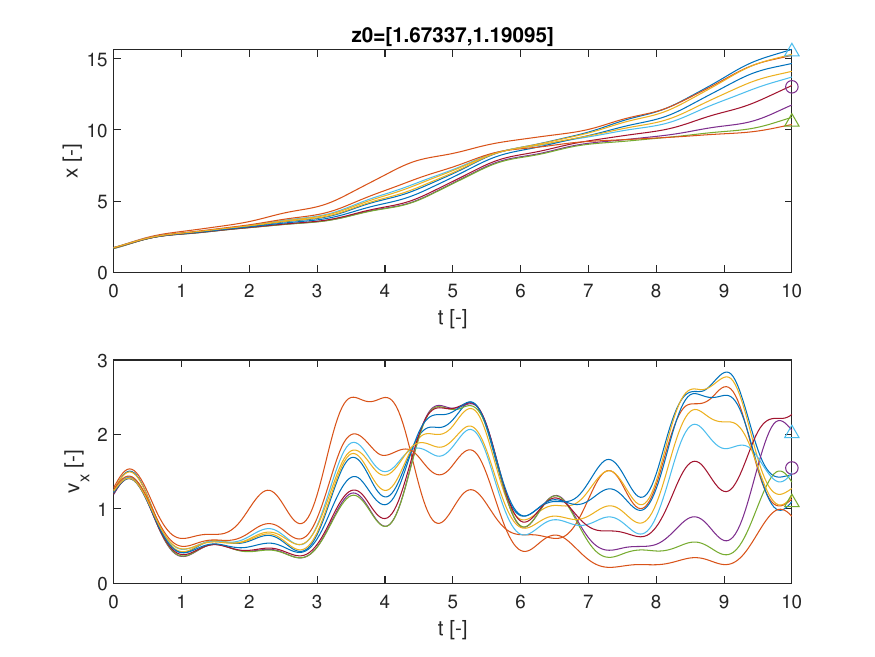}}}\\
     \caption{Two example of trajectory ensembles: a) low $\tilde{\alpha}$, b) high $\tilde{\alpha}$}
     \label{fig:pendulum_trj}
\end{figure}

Finally Figure \ref{fig:pendulum_trj} shows two notable trajectories, one for initial conditions $\mathbf{z}_0=[0.889447,\,-0.19598]$ and the other for $\mathbf{z}_0=[1.67337,1.19095]$, which correspond respectively to low and high values of $\tilde{\alpha}$. In this case 10 trajectories were propagated for 10 random realisations of $a$.

\subsection{The Uncertain Double Gyre}
The double gyre model consists of a pair of counter-rotating gyres, with a time-periodic perturbation. The system is modelled as a first order system of differential equations, given by:

\begin{equation}\label{eq:dgyre}
\dot{\mathbf{z}} = \frac{\drv}{\drv t}
    \begin{array}{c}
\begin{bmatrix}
x \\
y
\end{bmatrix}
\end{array}
=
\pi A 
\begin{bmatrix}
- \sin(\pi f(x,t))\cos(\pi y) \\
\cos(\pi f(x, t)) \sin(\pi y) \frac{\partial f}{\partial x}
\end{bmatrix}
=
\mathbf{g}(\mathbf{z}, \mathbf{p}, t)
\end{equation}
The functions and the coefficients appearing in the dynamics are given by:
\begin{equation}
\label{eq:gyreparam}
\begin{array}{l}
f(x, t) = a(t) x^2 + b(t) x \\
a(t) = \eta \sin(\omega t) \\
b(t) = 1 - 2 \eta \sin(\omega t) \\
A = 0.1, \ \ \ \ \omega = 2 \pi / 10 \\
\end{array}
\end{equation}

The Jacobian of the velocity field is given by:
\begin{equation}
    \label{eq:gyrejacob}
    \frac{\partial \textbf{g}}{\partial \textbf{z}}
=
\pi A
\begin{array}{cc}
\begin{bmatrix}
- \pi \cos(\pi y) \cos(\pi f) \frac{\partial f}{\partial x}, & \sin(\pi f) \sin(\pi y) \\
-\pi \sin(\pi f) \frac{\partial f}{\partial x}^2 + 2 a(t) \cos(\pi f), & \pi \cos(\pi f) \frac{\partial f}{\partial x} \cos(\pi y)
\end{bmatrix}
\end{array}
\end{equation}

We generalise results from previous works (e.g., \cite{faraz2012}), by considering the uncertain parameter $p=\eta$ to be an uncertain parameter defined over the interval $\eta\in[0.1-0.01,0.1+0.01]$.

\begin{figure}[htb]
    \centering
     \begin{subfigure}[b]{0.49\textwidth}
         \centering
         \includegraphics[width=1.1\textwidth]{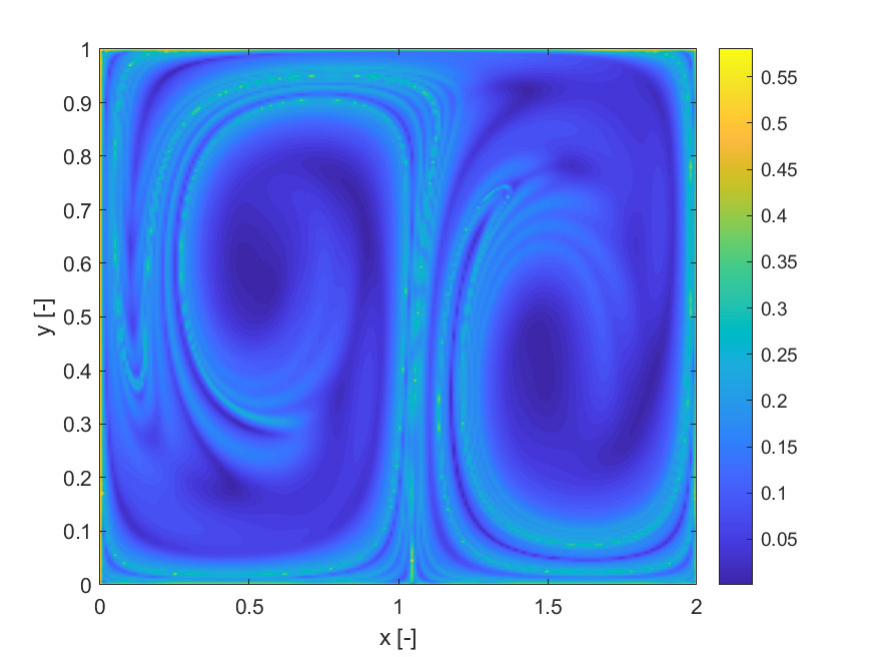}
         \caption{FTLE}
     \end{subfigure}    
     \begin{subfigure}[b]{0.49\textwidth}
         \centering
         \includegraphics[width=1.1\textwidth]{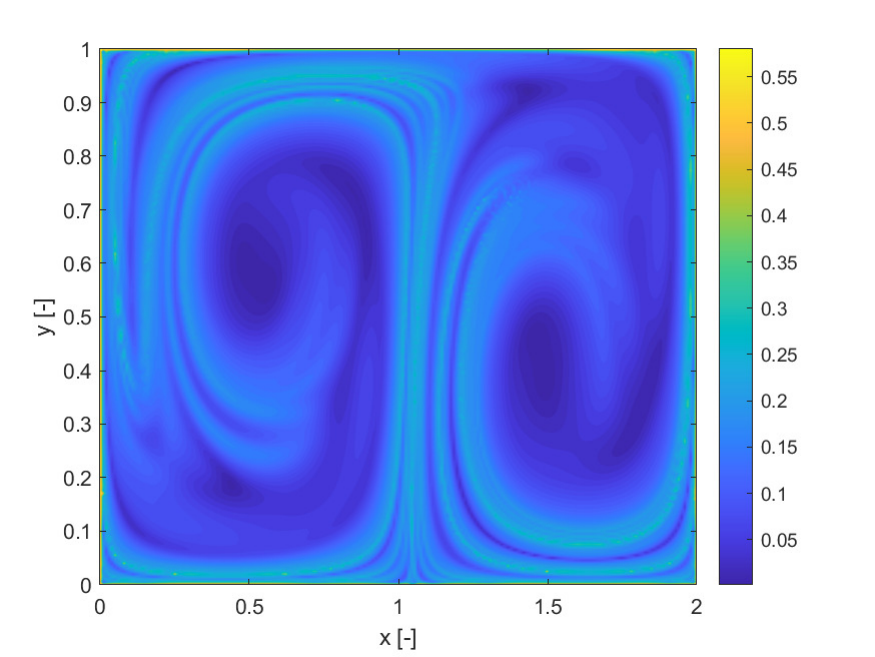}
         \caption{$\alpha_1^1$}
     \end{subfigure}
     \hfill
     \begin{subfigure}[b]{0.49\textwidth}
         \centering
         \includegraphics[width=1.1\textwidth]{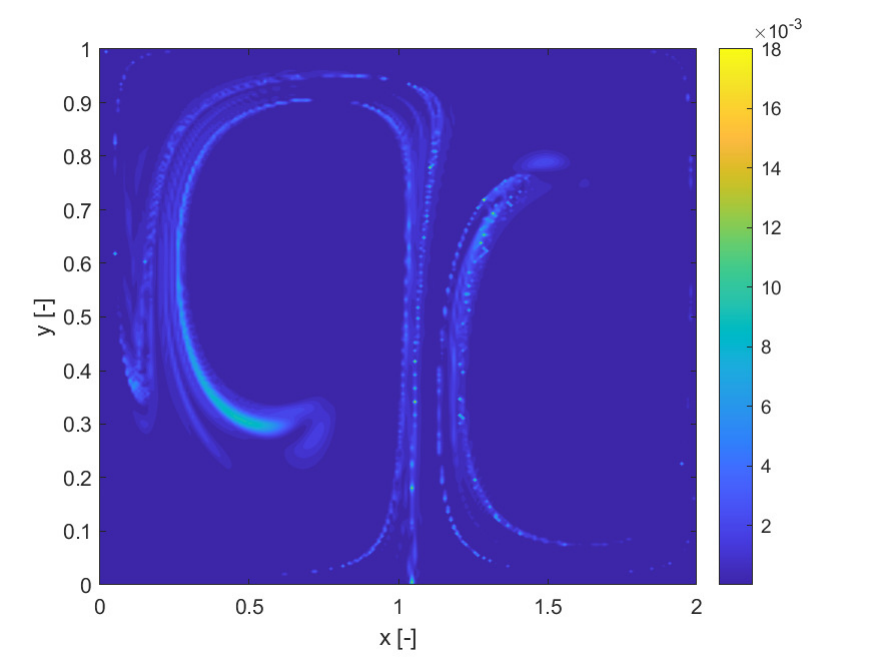}
         \caption{$\alpha_1^2$}
     \end{subfigure}
     \begin{subfigure}[b]{0.49\textwidth}
         \centering
         \includegraphics[width=1.1\textwidth]{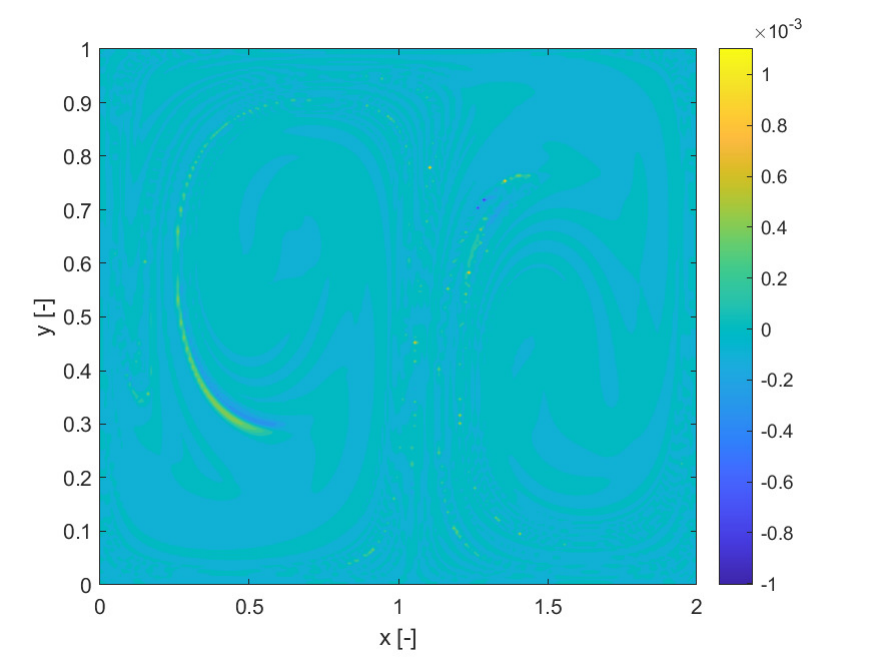}
         \caption{$\alpha_1^3$}
     \end{subfigure}
         \hfill
        \caption{SFTLE1, scalar fields of the double-gyre model. Integration time is $t_f = 20$.}
        \label{fig:SFTLE1_dgyre}
\end{figure}

\begin{figure}[htb]
    \centering
     \begin{subfigure}[b]{0.49\textwidth}
         \centering
         \includegraphics[width=1.1\textwidth]{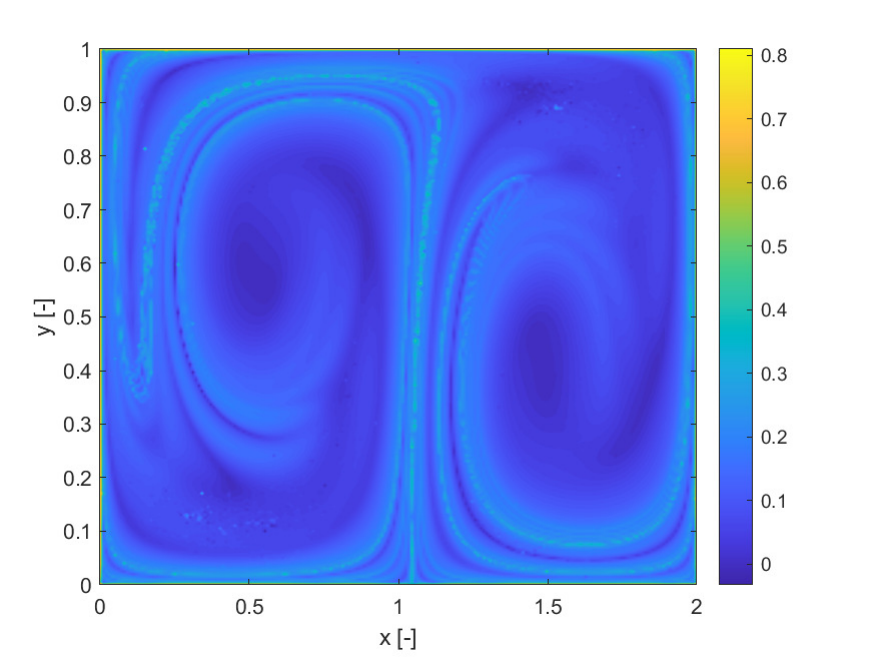}
         \caption{$\alpha_2^1$}
     \end{subfigure}
     \hfill
     \begin{subfigure}[b]{0.49\textwidth}
         \centering
         \includegraphics[width=1.1\textwidth]{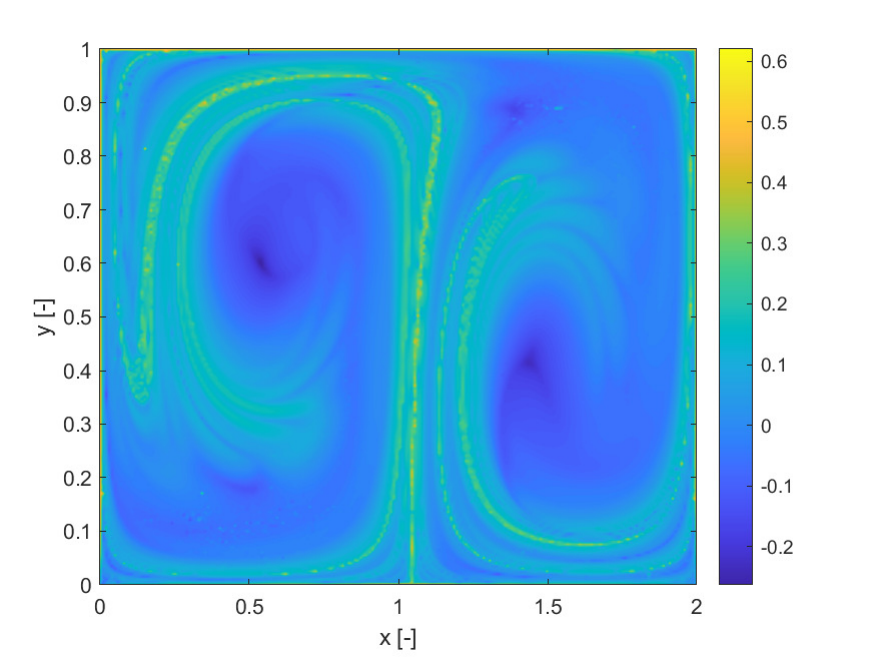}
         \caption{$\alpha_2^2$}
     \end{subfigure}
     \hfill
     \begin{subfigure}[b]{0.49\textwidth}
         \centering
         \includegraphics[width=1.1\textwidth]{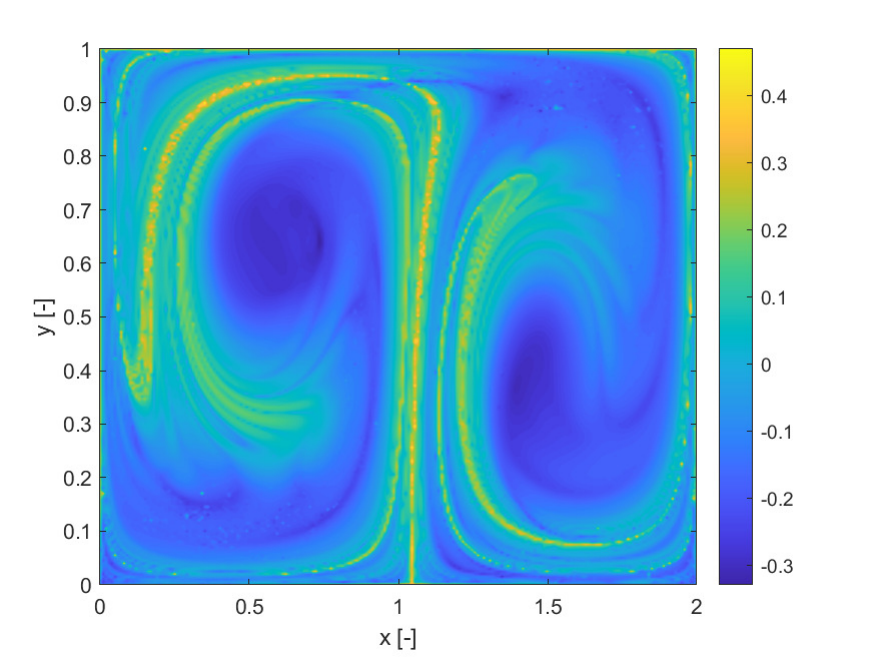}
         \caption{$\alpha_2^3$}
     \end{subfigure}     
        \caption{SFTLE2 scalar fields of the double-gyre model. Integration time is $t_f = 20$.}
        \label{fig:SFTLE2_dgyre}
\end{figure}

 \begin{figure}[htb]
     \begin{subfigure}[b]{0.53\textwidth}
     \centering
     {\includegraphics[width=1.0\textwidth]{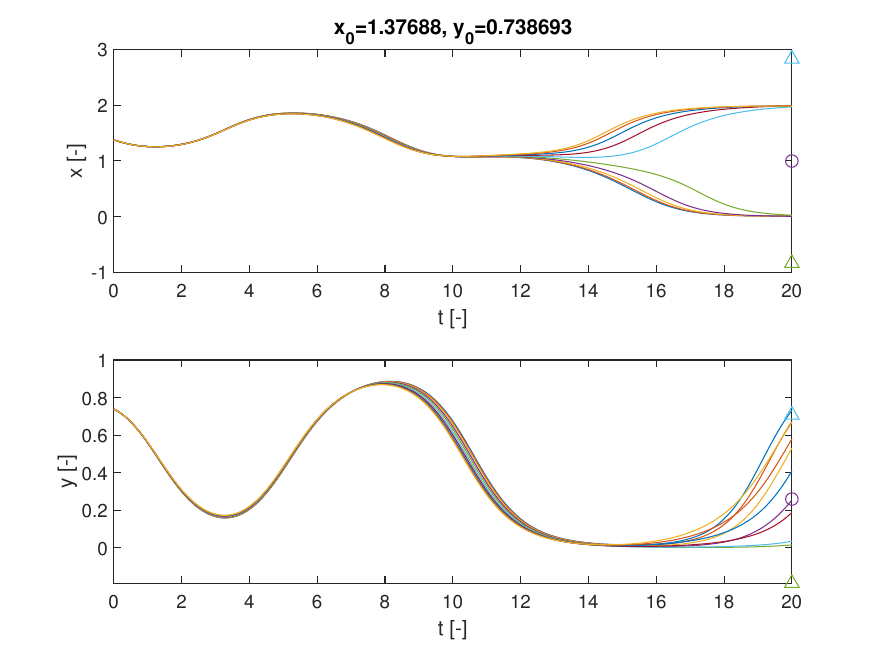}}\\
     \caption{High value of $\tilde{\alpha}$}
     \end{subfigure} 
\begin{subfigure}[b]{0.53\textwidth}
     \centering
     {\includegraphics[width=1.0\textwidth]{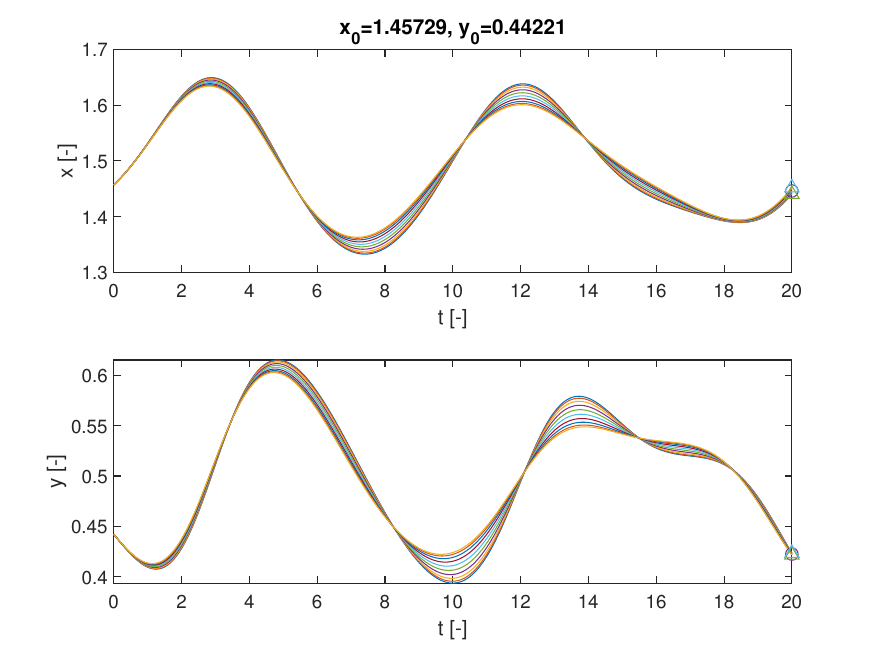}}\\
     \caption{Low value of $\tilde{\alpha}$}
     \end{subfigure}
     \caption{Examples of trajectory ensembles for high and low values of $\tilde{\alpha}$. Double gyre model.}        \label{fig:gyre_trj}
\end{figure}

 \begin{figure}[htb]
  \centering
     \begin{subfigure}[b]{0.49\textwidth}
     \centering
     {\includegraphics[width=1.1\textwidth]{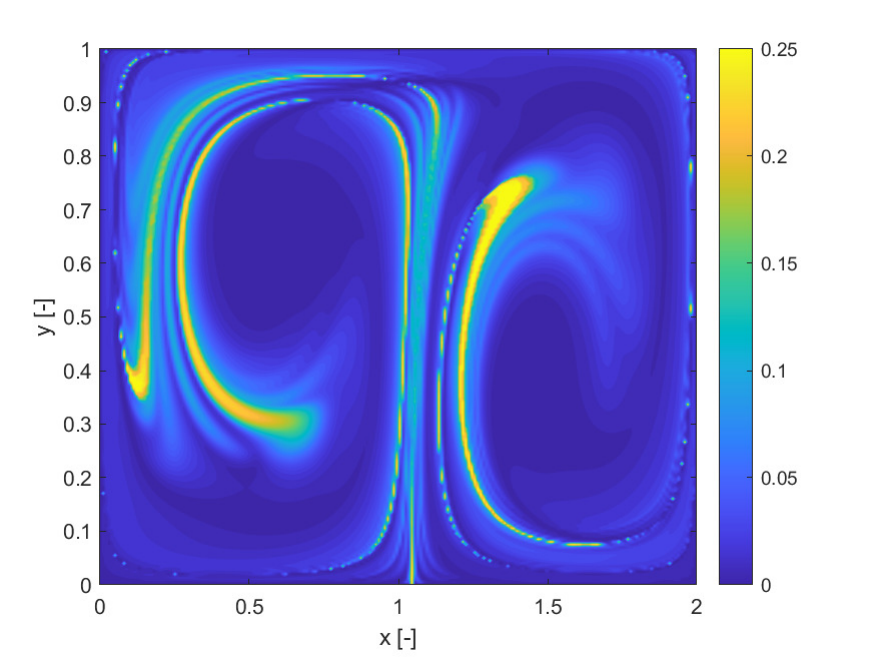}}\\
     \caption{$\tilde{\alpha}$}
     \end{subfigure} 
\begin{subfigure}[b]{0.49\textwidth}
     \centering
     {\includegraphics[width=1.1\textwidth]{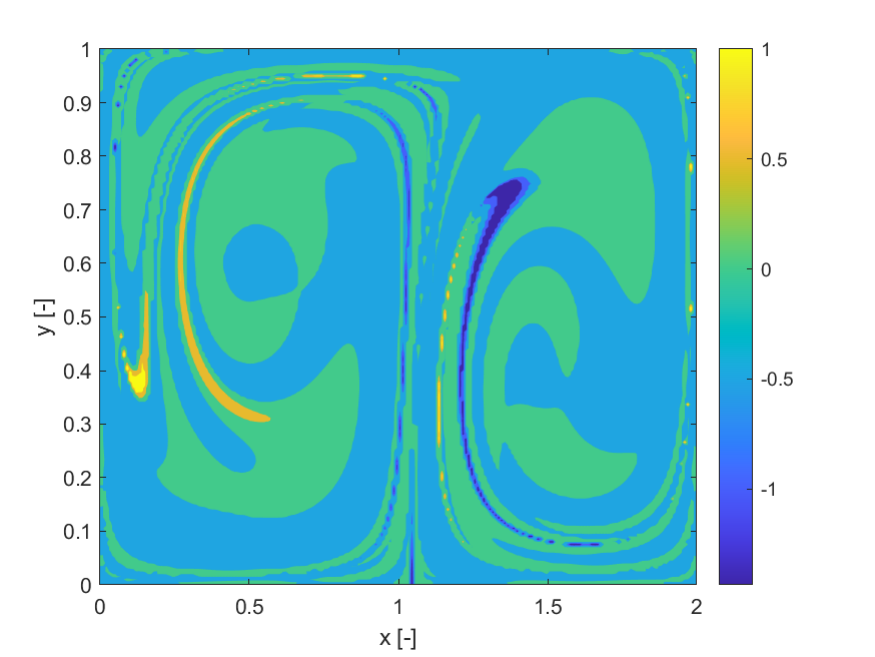}}\\
     \caption{Skewness of the $x$ component}
     \end{subfigure}
\begin{subfigure}[b]{0.49\textwidth}
     \centering
     {\includegraphics[width=1.1\textwidth]{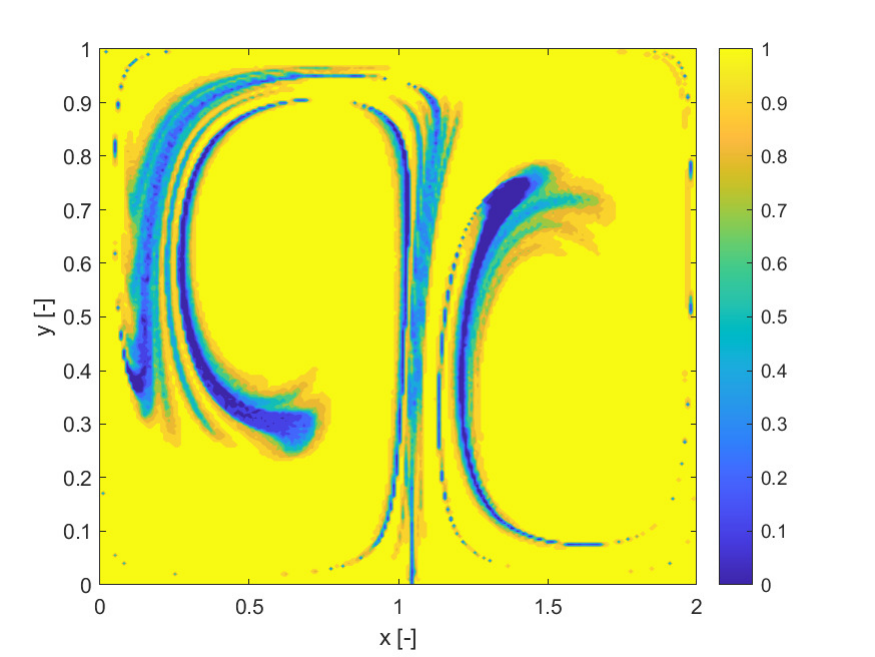}}\\
     \caption{$\mathbb{E}_{0.25}$}
     \end{subfigure}     
     \caption{Pseudo diffusion exponent for the double gyre model.}     \label{fig:sdi_gyre}     
\end{figure}


As in the previous example we expand the state variables in Chebyshev polynomials of degree 4. Differential equations are propagated with the same adaptive Runge-Kutta integrator with the same absolute and relative tolerances.
The propagation is performed for a fixed integration time $t_f = 20$.
A uniform grid of initial conditions has a size of $200\times200$, in the domains $x \in [0, 2]$, $y \in [0, 1]$. The finite increment for the calculation of both the FTLE and SFTLE is $\Delta z_j = 1 \cdot 10^{-7}$.

Figures \ref{fig:SFTLE1_dgyre}a and b compare the deterministic FTLE with SFTLE1. Also in this case $\alpha_1^1$ shows the same structures as the FTLE. It is interesting to note in Figure \ref{fig:SFTLE1_dgyre}c, how the location of the ridges of $\alpha_1^2$, are located near the ridges of $\alpha_1^1$. This implies that for chaotic initial conditions the set of trajectories behaves qualitatively differently with different realisations of the uncertain parameter. This emerges also from \ref{fig:SFTLE1_dgyre}d where the skewness of FLTE is positive or negative depending on the initial conditions.
Similar consideration can be derived from Figure \ref{fig:SFTLE2_dgyre} where the SFTLE2 are represented and from Figure \ref{fig:sdi_gyre} where $\tilde{\alpha}$ is represented together with the expectation for a threshold of $\epsilon=0.25$ and the skewness of the x component of the ensemble of trajectories.Ten trajectories corresponding to ten realisation of $\eta$ are represented in Figure \ref{fig:gyre_trj} for two initial conditions $x_0=1.37688$, $y_0=0.73869$ and $x_0=1.45729$, $y_0=0.44221$ corresponding respectively to high and low values of $\tilde{\alpha}$. Note in Figure \ref{fig:gyre_trj}a the bifurcation of the ensemble into two different groups of trajectories.

\subsection{The Uncertain Circular Restricted Three-Body Problem} \label{sec:CR3BP}
The Circular Restricted Three-Body Problem (CR3BP) is arguably one of the most studied problems in celestial mechanics. In this section we will consider the planar case with an uncertain mass parameter. The planar circular restricted three-body problem \cite{szeb} is governed by:

\begin{equation}
    \label{eq:3bp}
    \Ddot{x} - 2 \dot{y} = \frac{\partial J}{\partial x}
\end{equation}
$$
\Ddot{y} + 2 \dot{x} = \frac{\partial J}{\partial y}
$$

where $J(x,y)$ is given by:

\begin{equation}
\label{eq:omega}
    J(x,y) = \frac{x^2+y^2}{2} + \frac{1 - \mu}{\sqrt{(x+ \mu)^2 + y^2 }} + \frac{\mu}{\sqrt{(x - 1 + \mu)^2 + y^2}} + \frac{1}{2} \mu (1 - \mu)
\end{equation}

and $\mu$, the mass parameter of the system, is a function of the masses of the primaries. With this formulation, the reference frame is uniformly rotating and the primaries' position, in such frame, is constant in time.
We can again re-write the system as a first order system of differential equations:

\begin{equation}
    \label{eq:3bp1ode}
        \dot{\mathbf{z}} = \frac{\drv}{\drv t}
    \begin{array}{c}
\begin{bmatrix}
x \\
y \\
v_x \\
v_y
\end{bmatrix}
\end{array}
=
    \begin{array}{c}
\begin{bmatrix}
v_x \\
v_y \\
2 v_y + \frac{\partial J}{\partial x} \\
- 2 v_x + \frac{\partial J}{\partial y}
\end{bmatrix}
\end{array}
=
\mathbf{g}(\mathbf{z}, p)
\end{equation}
with $v_x=\dot{x}$ and $v_y=\dot{y}$ and uncertain parameter $p=\mu$. The partial derivatives of $J$, appearing in \eqref{eq:3bp1ode}, are given by:

\begin{equation}
\label{eq:partialOmega}
    \frac{\partial J}{\partial x} = x - \frac{(1 - \mu)(x+ \mu)}{((x+ \mu)^2 + y^2 )^{3/2}} - \frac{\mu (x - 1 + \mu)}{((x - 1 + \mu)^2 + y^2)^{3/2}}
\end{equation}
$$
\frac{\partial J}{\partial y} = y - \frac{y(1 - \mu)}{((x+ \mu)^2 + y^2 )^{3/2}} - \frac{\mu y}{((x - 1 + \mu)^2 + y^2)^{3/2}}
$$

The Jacobian of the velocity field, associated to the first-order formulation of the dynamics, is:

\begin{equation}
    \label{eq:var3bp}
    \frac{\partial \textbf{g}}{\partial \textbf{z}}
=
\begin{array}{cccc}
\begin{bmatrix}
0 & 0 & 1 & 0 \\
0 & 0 & 0 & 1 \\
\frac{\partial^2 J}{\partial x^2} & \frac{\partial^2 J}{\partial y \partial x} & 0 & 2 \\
\frac{\partial^2 J}{\partial x \partial y} & \frac{\partial^2 J}{\partial y^2} & -2 & 0
\end{bmatrix}
\end{array}
\end{equation}

in which the second order derivatives of $J$ are not expanded for brevity. The energy is then defined as,

\begin{equation}
\label{eq:energy}
    E(x, y, v_x, v_y) = \frac{1}{2} (v_x^2 + v_y^2) - J(x, y) \end{equation}
    
and is a constant of motion for the CR3BP. In order to reduce the dimensionality of the problem from four to two, the initial conditions are defined as:

\begin{equation}
    \textbf{z}_i = [x_i, 0, v_{xi}, v_{y}(x_i, 0, v_{xi}, E_0)]
\end{equation}

where the value of $v_y$ is computed, from a given condition $(x_i, v_{xi})$, making use of the conservation of energy:

\begin{equation}
    \label{eq:ydot}
    v_y = - \sqrt{2(E_0 + J) - v_x^2}
\end{equation}
For the results in this paper, the energy level has been set to $E_0= E(L_1) + 0.03715$, where $E(L_1) = E(L_1^x, 0, 0, 0)$ is the potential energy at the first Lagrange point, $L_1^x$ being given \cite{wakker}by:
\begin{equation}
    \label{eq:L1x}
    L_1^x = 1 - \mu - \gamma_1
\end{equation}
with
$$
\gamma_1 = b - \frac{1}{3} b^2- \frac{1}{9} b^3 - \frac{23}{81} b^4
$$
and
$$
b = \left( \frac{1}{3} a \right)^{1/3};\;\; a = \frac{\mu}{1 - \mu}
$$



\begin{figure}[htb]
    \centering
     \begin{subfigure}[b]{0.45\textwidth}
         \centering
         \includegraphics[width=\textwidth]{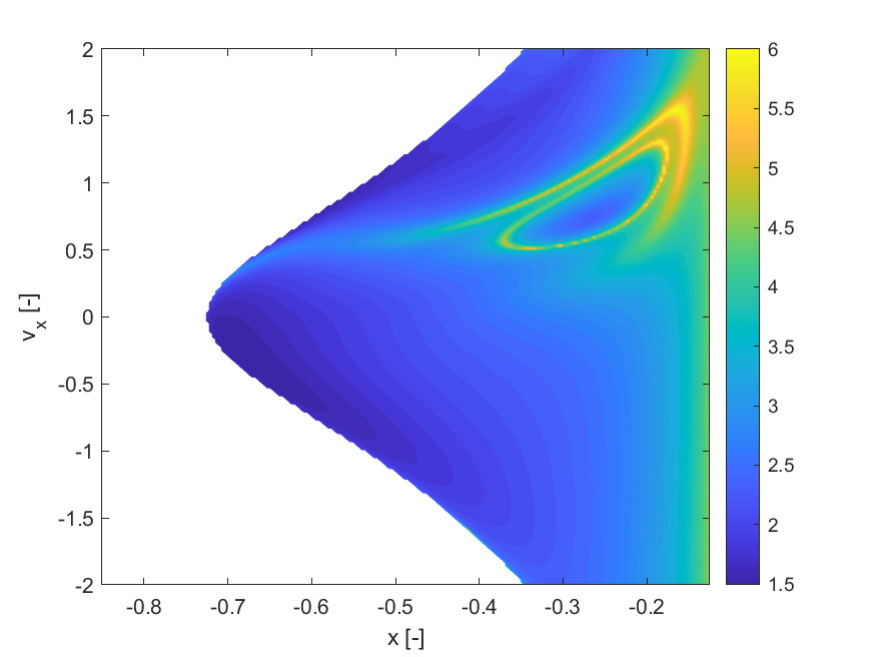}
         \caption{FTLE}
     \end{subfigure}
     \hfill
     \begin{subfigure}[b]{0.45\textwidth}
         \centering
         \includegraphics[width=\textwidth]{3bp_moments/ftle_3bp_T20_smooth.pdf}
         \caption{$\alpha_1^1$}
     \end{subfigure}
     \begin{subfigure}[b]{0.45\textwidth}
         \centering
         \includegraphics[width=\textwidth]{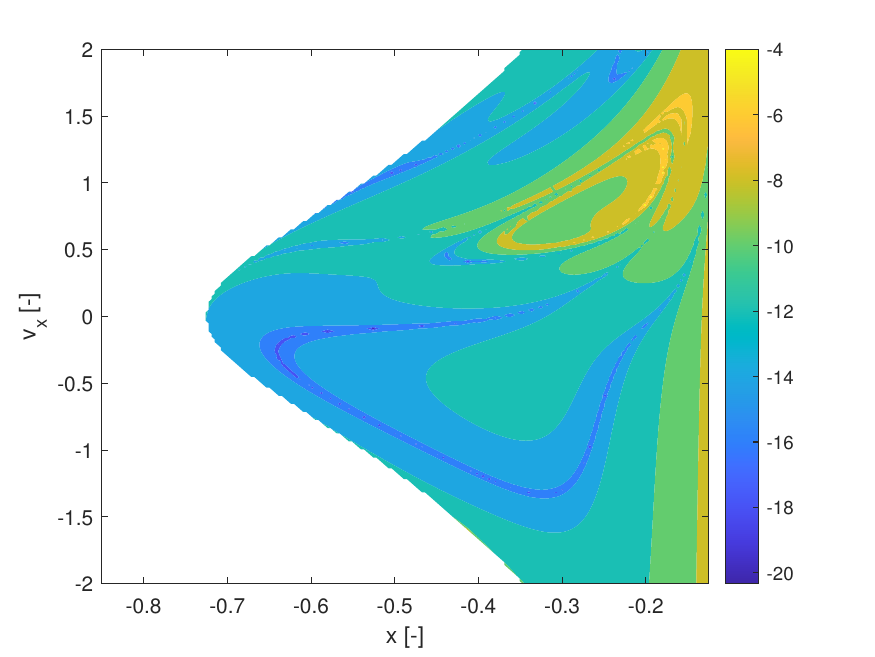}
         \caption{$\log_{10}\alpha_1^2$}
     \end{subfigure}
         \hfill
     \begin{subfigure}[b]{0.45\textwidth}
         \centering
         \includegraphics[width=\textwidth]{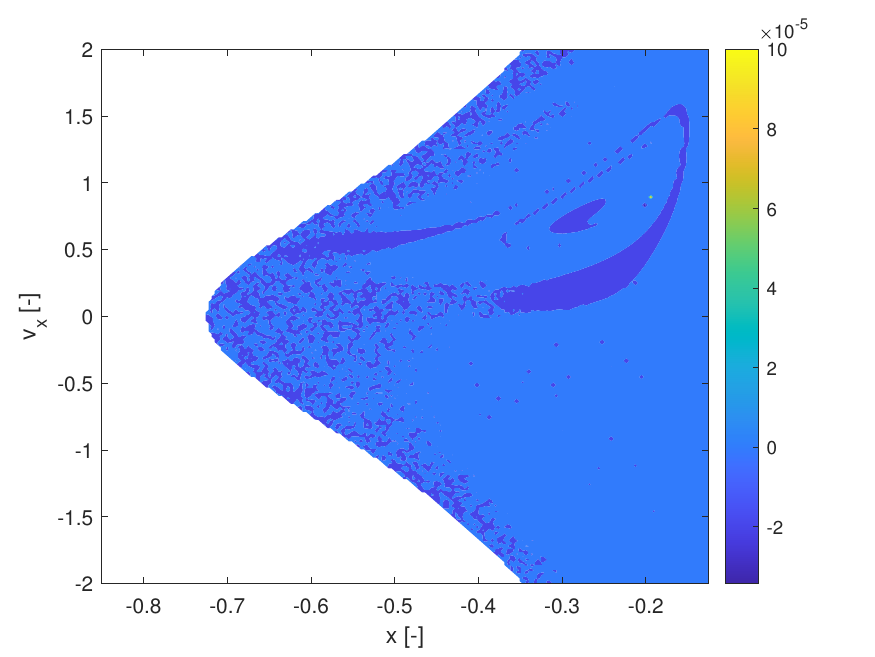}
         \caption{$\alpha_1^3$}
     \end{subfigure}
        \caption{FTLE and SFTLE1 scalar fields for the CR3BP model case 1. Integration time is $t_f = 2$.}
        \label{fig:sftlecampagnola}
\end{figure}

We consider two cases. In case 1, the integration is performed for two full revolutions of the primaries, or $t_f = 2$, using an explicit adaptive Runge-Kutta 4/5 method with absolute tolerance $10^{-10}$ and relative tolerance of $10^{-8}$. As in the previous two examples, we use Chebyshev orthogonal polynomials of type 2, thus the integration abscissas and weights are optimised for these polynomials.
The initial conditions grid is $200\times200$, in the domains $x \in [-0.85, -0.125]$, $v_x \in [-2.0, 2.0]$. The value of the mass parameter is assumed to be uncertain and within the interval reported in Table \ref{tab:3bp_param} case 1.
The finite increment for the calculation of both the FTLE and SFTLE is $\Delta z_j = 1 \cdot 10^{-7}$. Figure \ref{fig:sftlecampagnola} reports the FTLE and SFTLE1 for case 1. Polynomials are expanded to order 4 and the figure represents the first three SFTLE1.

\begin{table}[]
    \centering
        \caption{Summary of parameter settings for the 2 cases of the uncertain CR3BP}
    \begin{tabular}{|c|c|c|}
         \hline
         Case & Mass Parameter & Integration Time\\
         \hline
         1 & $\mu\in [0.1-10^{-7},1+10^{-7}]$& $t_f=2$\\  
         \hline
         2 & $\mu\in [0.1-10^{-3},1+10^{-3}]$& $t_f=2.8$\\
         \hline
    \end{tabular}
    \label{tab:3bp_param}
\end{table}

\begin{figure}[htb]
    \centering
     \begin{subfigure}[b]{0.47\textwidth}
         \centering
         \includegraphics[width=1.1\textwidth]{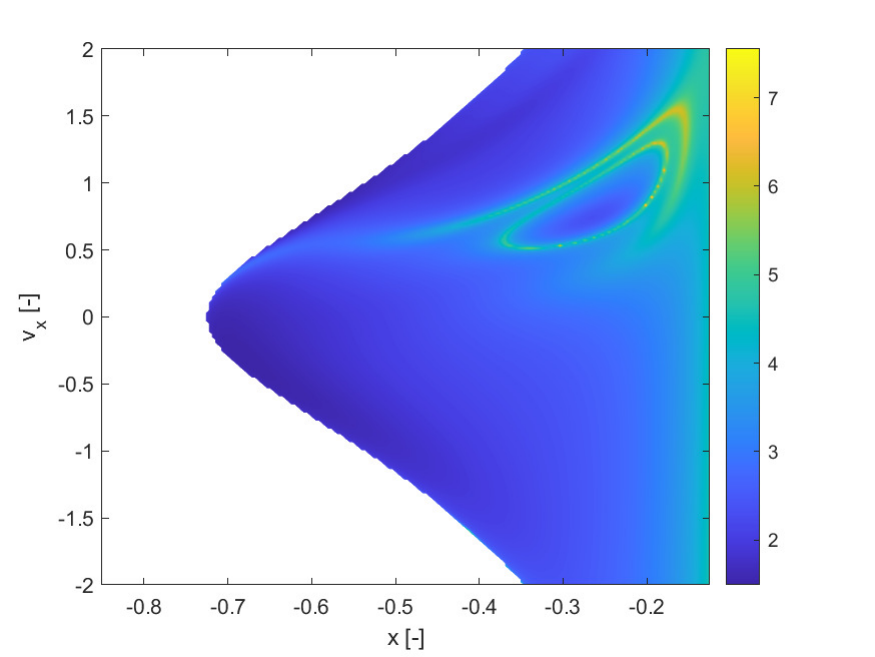}
         \caption{$\alpha_2^1$}
     \end{subfigure}
     \begin{subfigure}[b]{0.47\textwidth}
         \centering
         \includegraphics[width=1.1\textwidth]{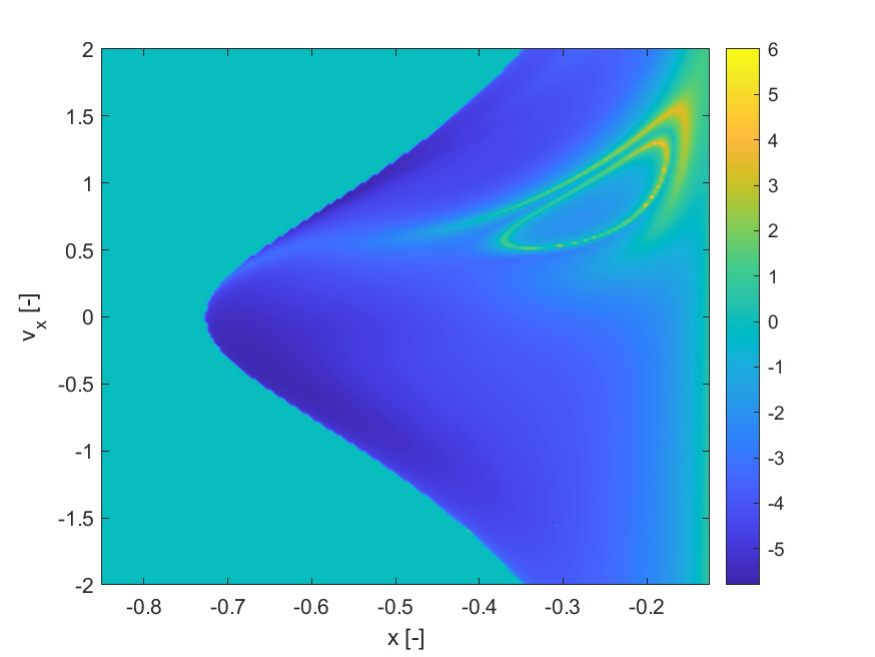}
         \caption{$\alpha_2^2$}
     \end{subfigure}
         \hfill
     \begin{subfigure}[b]{0.47\textwidth}
         \centering
         \includegraphics[width=1.1\textwidth]{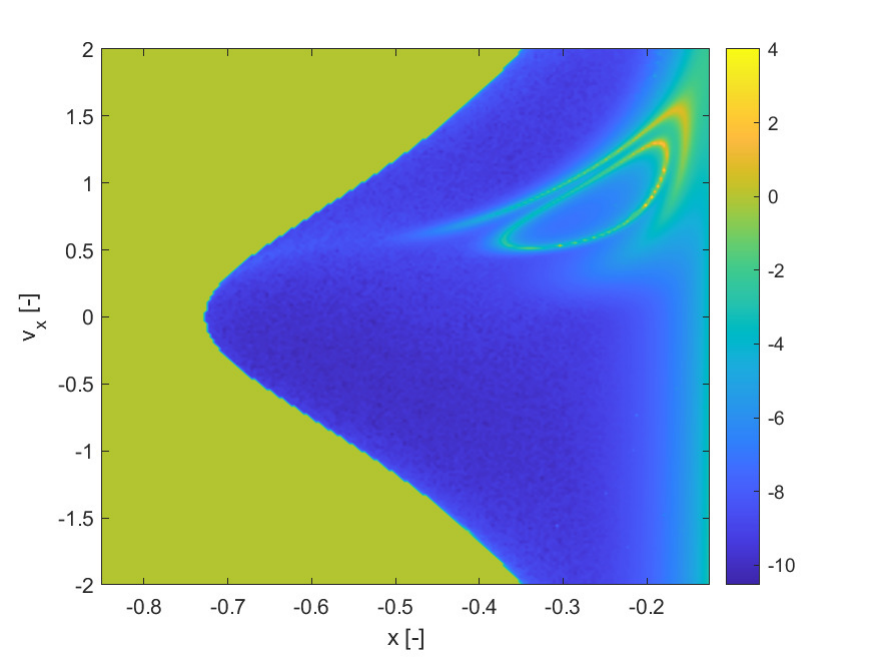}
         \caption{$\alpha_2^3$}
     \end{subfigure}
        \caption{SFTLE2 scalar fields for the CR3BP model case 1. Integration time is $t_f = 2$.}
        \label{fig:sftl2}
\end{figure}

In Figures \ref{fig:sftlecampagnola}a, b and c, the intersection of the invariant manifold with the plane $y=0$ is identified by the closed yellow ridge in the upper right part of the figures. As it was found also in previous works, the presence of ridges in FTLE fields is only a sufficient condition for the existence of Lagrangian Coherent Structures (and invariant manifolds in particular), but not a necessary one. In fact, other ridges in the same figures are not associated to manifold crossings. Figure \ref{fig:sftl2} shows the first three SFTLE2 for the same case. Also in this case the ridges are consistent with the ones in the FTLE plot and the range of the indicator is progressively shifted towards negative values.

\begin{figure}[htb]
    \centering
         \includegraphics[width=0.8\textwidth]{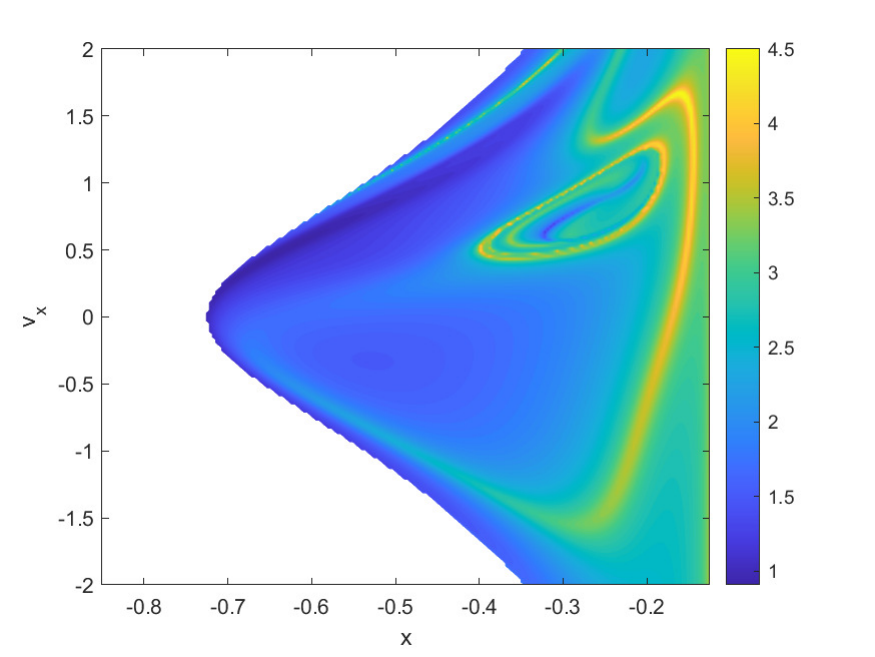}
         \caption{FTLE field of the CR3BP for an integration time $t_f=2.8$}
\label{fig:ftle28}         
\end{figure}

\begin{figure}[htb]
     \centering
     \subfloat[$\tilde{\alpha}$]{\includegraphics[width=0.5\textwidth]{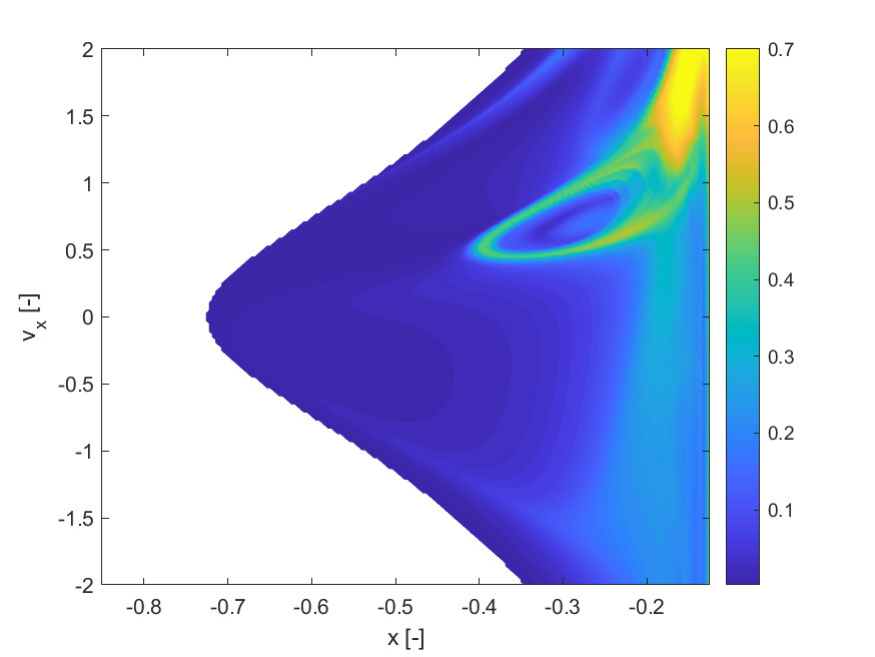}}
     \subfloat[$\mathbb{E}_{0.1}$]{\includegraphics[width=0.5\textwidth]{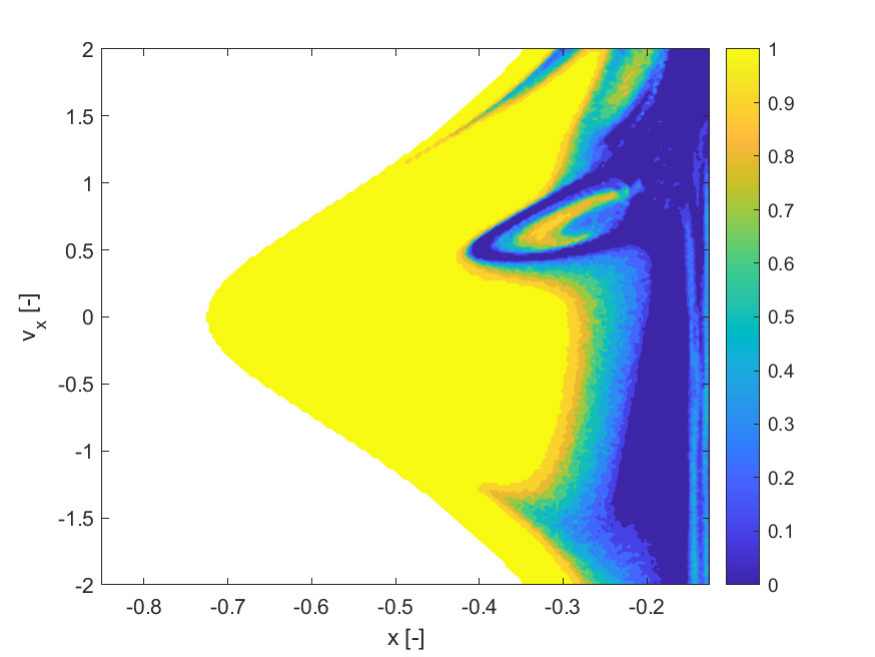}}\\
     \caption{Pseudo-diffusion exponent for the CR3BP case 2.}
     \label{fig:3bp_alpha}
\end{figure}

For case 2, we extended the integration time and also the range of the uncertain parameter (see case 2 in Table \ref{tab:3bp_param}). The extension of the integration time allows one to observe more interesting behaviours. In particular some trajectories start from the primary with coordinate $x=1-\mu$ and flow to the primary with coordinate $x=\mu$. Figure \ref{fig:ftle28} shows the FTLE field for case 2. For this second case we build a cartography only with the pseudo diffusion exponent $\tilde{\alpha}$ because it was faster than the computation of SFTLE1 and 2 and gave interesting results. Figure \ref{fig:3bp_alpha} shows the $\tilde{\alpha}$ field for the CR3BP case 2 together with the expectation $\mathbb{E}$ for a threshold $\epsilon=0.1$.

\begin{figure}[htb]
     \centering
     \subfloat[$\log_{10}\tilde{\alpha}_x$]{{\includegraphics[width=0.5\textwidth]{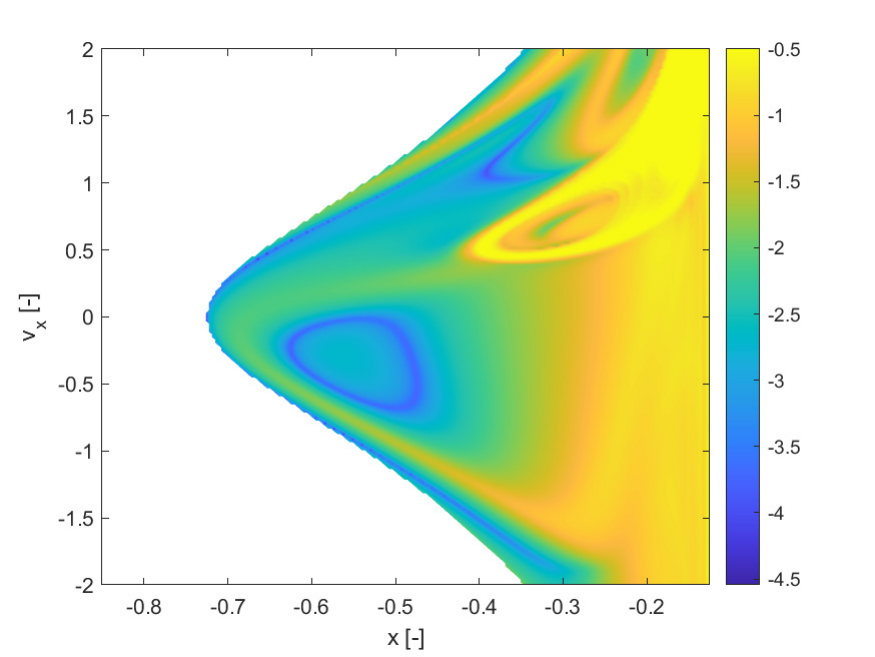}}}
     \subfloat[$\log_{10}\tilde{\alpha}_y$]{{\includegraphics[width=0.5\textwidth]{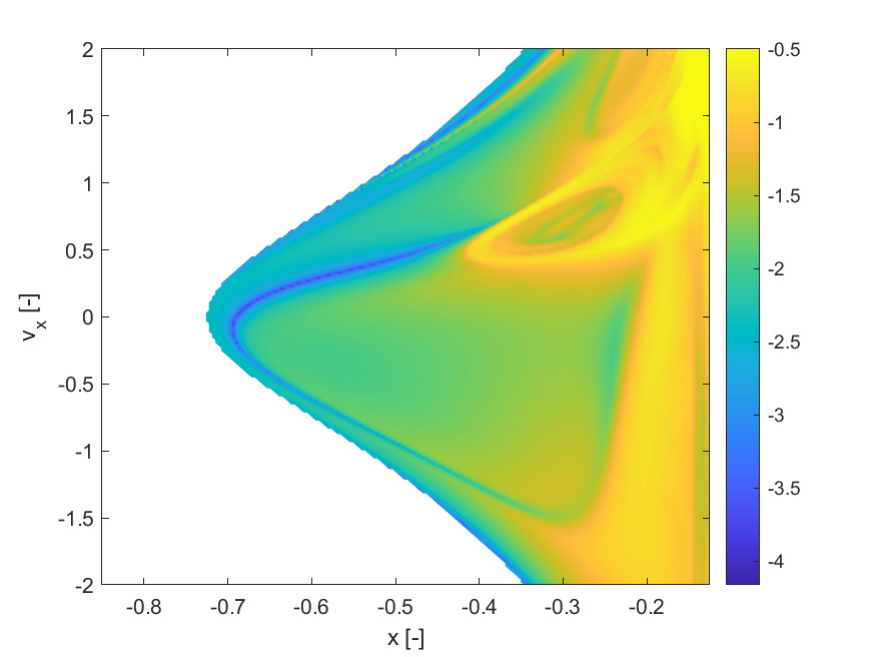}}}\\
     \caption{Pseudo-diffusion exponent for the Cr3BP case 2: individual components.}
     \label{fig:3bpxy_alpha}
\end{figure}

\begin{figure}[htb]
     \centering
     \subfloat[]{{\includegraphics[width=0.5\textwidth]{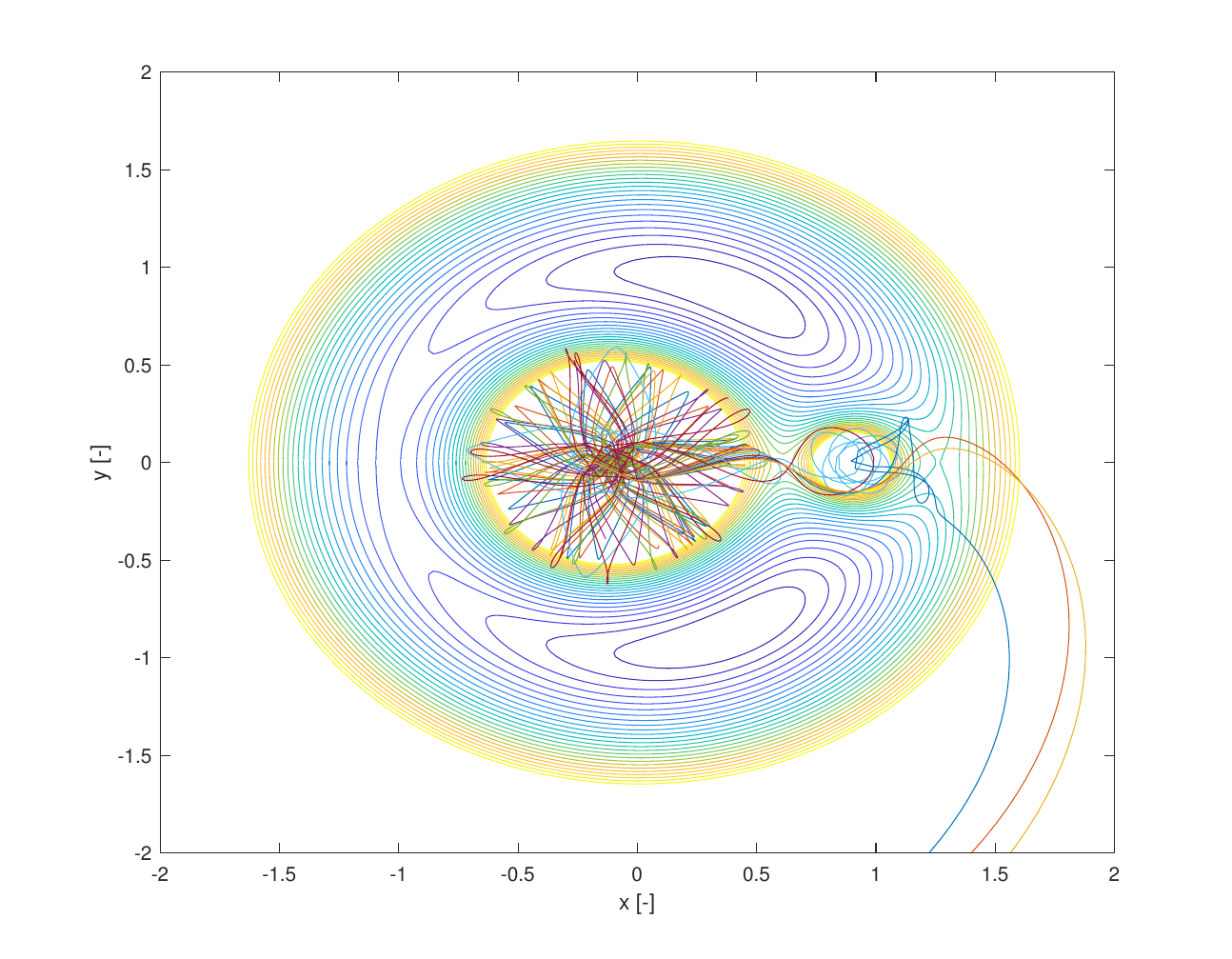}}}
     \subfloat[]{{\includegraphics[width=0.5\textwidth]{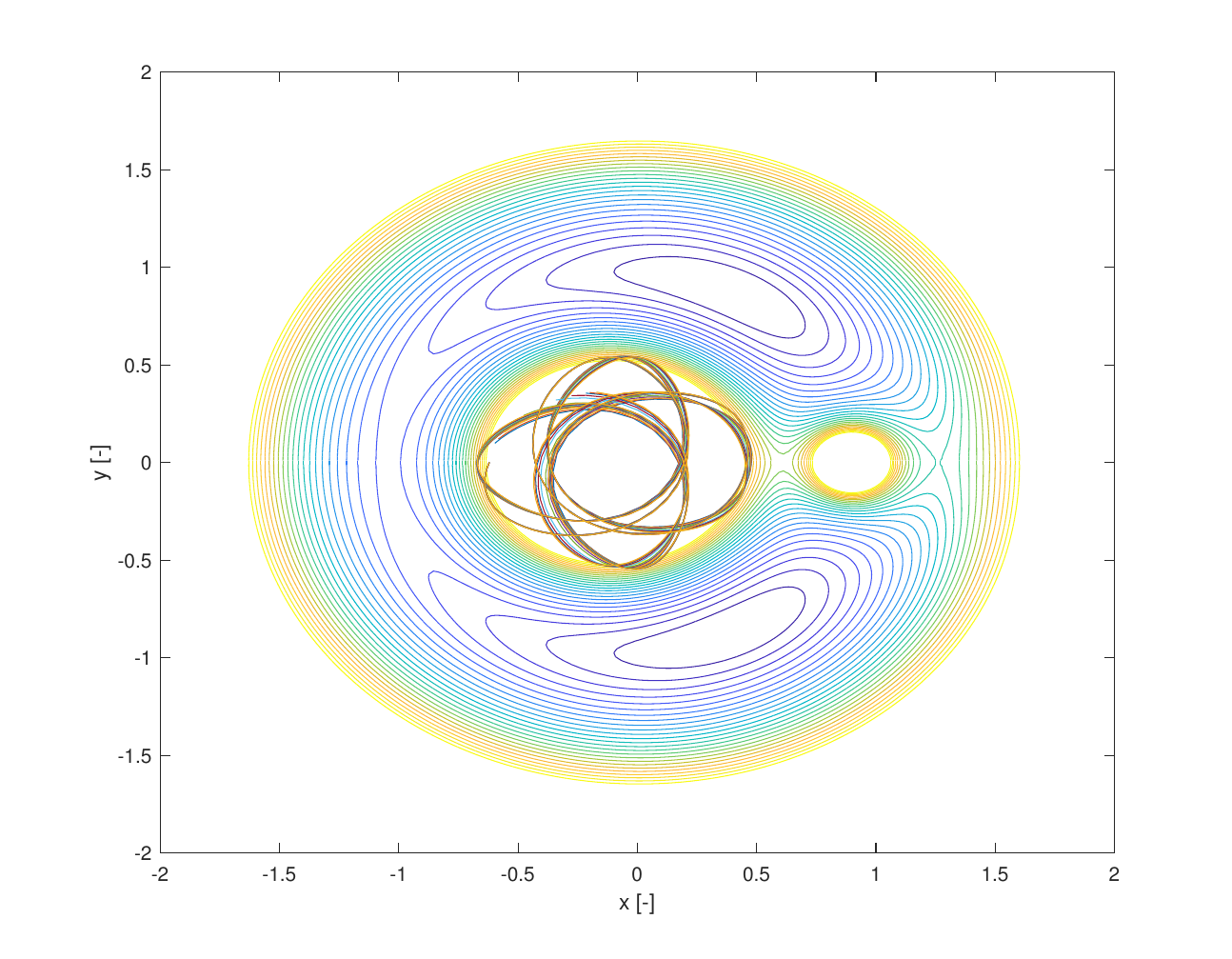}}}\\
     \caption{Example of ensemble of trajectories for: a) highly diffusive case, b) very low pseudo diffusion exponent. Integration time $t_f=28$.}
     \label{fig:3bp_trj_alpha}
\end{figure}

Figure \ref{fig:3bpxy_alpha} shows the $\tilde{\alpha}_x$ and $\tilde{\alpha}_y$ fields respectively. Figure \ref{fig:3bp_trj_alpha} presents two trajectory ensembles for two extreme cases of very low and very high values of $\tilde{\alpha}$ propagated for a time $t_f=28$. In particular, Figure \ref{fig:3bp_trj_alpha}a corresponds to initial conditions $x_0=-0.157789, v_{x0}=1.63819$ and Figure \ref{fig:3bp_trj_alpha}b corresponds to initial conditions $x_0=-0.624121, v_{x0}=-0.271357$. The latter corresponds to a point in the blue ring in Figure \ref{fig:3bpxy_alpha}a while the former corresponds to a point in the yellow region in Figure \ref{fig:3bp_alpha}.
The ensemble of trajectories is superimposed to the level curves of $J$ calculated with a fixed $\mu=0.1$ and we limit the axes for $x$ and $y$ to the interval $[-2,2]$ and $[-2,2]$. 

It is remarkable that $\tilde{\alpha}$ captures the diffusion of trajectories that eventually leave the system (see Figure \ref{fig:3bpxy_alpha}a) or trajectories that are quasi periodic (see Figure \ref{fig:3bpxy_alpha}b).
In the former case a change in the mass parameter, for a fixed value of the initial conditions, leads the total mechanical energy to fluctuate from a value below the zero velocity energy of the Lagrange equilibrium point 2 (L2) to one above it. Thus for some realisations of $\mu$ the zero velocity curves open at L2 and allow some trajectories to exit. In the latter case, instead all realisations remain confined and display very little sensitivity to the variation of $\mu$.

\section{Computational Complexity}
\label{sec:comp}
The computational cost of the SDIs is mainly dictated by the complexity of the calculation of the coefficients of the polynomial expansions.
The computational complexity of the pseudo diffusion exponent using non-intrusive polynomials requires the integration of $N$ sample trajectories. The number $N$ depends on the integration scheme. For a full tensor product and Gauss formulas $N=n_g^{n_p}$ with $n_g$ the number of integration points per uncertain dimension. For a sparse grid, the number of sample trajectories grows as $N=2^ll^{n_p-1}$ where $l$ is the level of the sparse grid. \textcolor{blue}{Thus in the examples presented above the pseudo diffusion exponent required the propagation of 9 trajectories per initial condition.} The number of coefficients to be computed for a full polynomial expansion is given by $M=\binom{n_p+m}{n_p}$ with a corresponding number of projection integrals.
If an intrusive method is used instead one needs to propagate $M$ differential equations and, for each equation, compute a multidimensional integral.

The computation of the SFTLE1 requires $N$ values of the FTLE. Since the computation of the FTLE requires propagating $2n$ tracers the computation of SFTLE1 requires $2Nn$ trajectories. \textcolor{blue}{In the test cases in the previous section, 9 Gauss integration points were used. Thus the computation of SFTLE1 required 36 propagations of the dynamics per initial condition for all two dimensional problems, and 72 propagations for the CR3BP.}
The computation of the SFTLE2, instead, requires the propagation of $2Mn$ equations and in each equation the dynamics is evaluated $N$ times per integration step. \textcolor{blue}{Looking at the examples in the previous sections, for an expansion to degree 3 and one uncertain parameter, the number of equations to propagate for each initial condition is 12, for a two dimensional problem, and 24, for a four dimensional problem, and for each equation the dynamics is evaluated 9 times per integration step.}
Thus in terms of number of propagation and computational cost the use of the pseudo diffusion exponent computed with non-intrusive expansions and sparse grids provides the fastest approach. 
If polynomials are propagated with an algebra the use of the SFTLE2 becomes an interesting option, along with the pseudo diffusion exponent, as it incorporates part of the sensitivity to the initial conditions. 

\section{Discussion}\label{sec:disc}
The three indicators proposed in this paper were shown to capture similar structures when applied to a cartographic study of dynamical systems under uncertainty. However, they measure conceptually different properties. SFTLE1 measures the statistical moments of the uncertainty in the standard FTLE. The first moment was shown to provide the same qualitative information of standard FTLE, while higher moments provide more interesting and unexpected information on the evolution of the dynamical system. In particular the strength of diffusive processes or asymmetries in the evolution of the system. SFTLE2 measures the divergence of neighbouring polynomial expansions. When this index is negative two polynomial expansions are behaving similarly up to time $t_f$. A value higher than zero means that the coefficients of the polynomials are diverging, which implies a divergent behaviour of the trajectories. Since the coefficients can be used to compute the statistical moments, divergent coefficients signifies that the ensemble of trajectories induced by multiple realisations of the uncertain parameters are also diverging.

In this sense analysing all the SFTLE2 with superscript up to $m$ might not bring additional useful information as the highest one is sufficient to understand the behaviour of the system. Thus one can argue that the maximal index $m$ of the positive SFTLE2 can work as a measure of the degree of divergence. This aspect needs further investigation before coming to a conclusion and will be the subject of future work. 

\begin{figure}[htb]
     \centering
     \subfloat[FTLE]{{\includegraphics[width=0.5\textwidth]{oscillator_moments/ftle_pendulum_smooth2.pdf}}}
     \subfloat[$\log_{10}\tilde{\alpha}$]{{\includegraphics[width=0.5\textwidth]{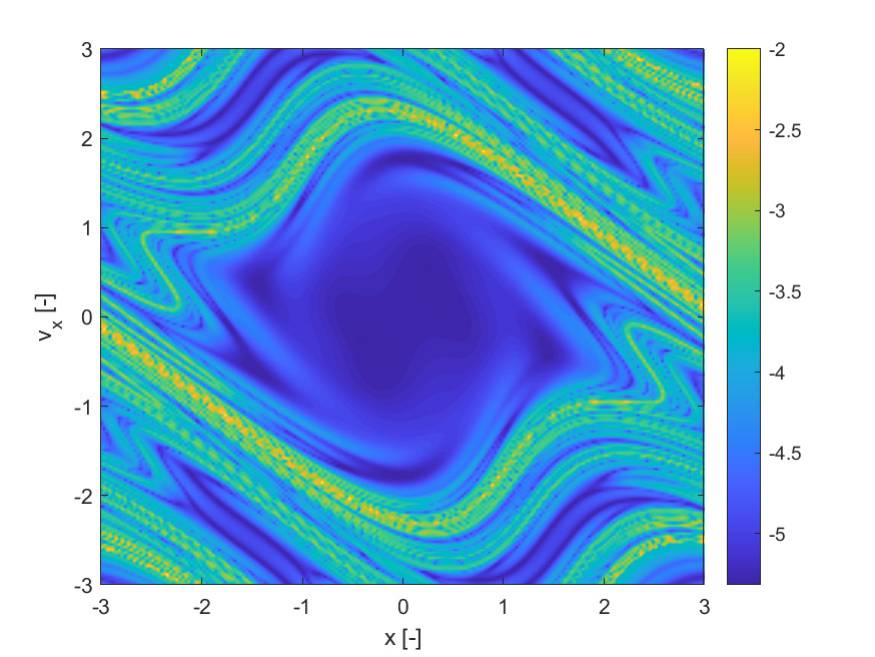}}}\\
     \caption{Uncertain pendulum. Comparison between pseudo diffusion exponent in the case of deterministic parameter $a$ and uncertain initial conditions and FTLE.}
     \label{fig:FTLE_vs_alpha}
\end{figure}

The pseudo diffusion exponent directly measures the degree of diffusion of the ensemble of trajectories. This suggests that the pseudo diffusion exponent of an infinitesimal uncertainty in the initial conditions would give the same qualitative information of the FTLE. This can be seen in Figure \ref{fig:FTLE_vs_alpha} where the FTLE for the uncertain perturbed pendulum is compared to the $\log10$ of $\tilde{\alpha}$. In this case $\tilde{\alpha}$ is computed with a simple first order polynomial expansion and only 9 integration points. The initial conditions are assumed to belong to a square with edge $10^{-5}$ centred in the nominal value of the initial conditions while the model parameter $a$ is deterministic and fixed at 2.5. Since the magnitude of the coefficients of the polynomial expansion is dependent on the magnitude of the uncertainty, an infinitesimal uncertainty leads to a small value of $\tilde{\alpha}$. However, from Figure \ref{fig:FTLE_vs_alpha} one can see a remarkable similitude between the FTLE and $\tilde{\alpha}$ to the point that the latter appears simply to be a scaled version of the former. 
This result can be understood if one considers the polynomial approximation of the propagated states. In fact assumes that one computed the FTLE from a linear approximation of $\mathbf{z}(t_f)$ with respect to the uncertain vector $\mathbf{p}=\mathbf{z}_0$ so that $\mathbf{z}(t_f)\approx \sum_i^m \mathbf{c}_{i}\Psi_i(\mathbf{p})$ with $m=1$, then we can demonstrate the following proposition. 
\begin{proposition}
\label{pro:initial_cond_alpha}
The eigenvalues $\lambda_i^{C_v}$ of the covariance matrix $\mathbf{C}_v$ in \eqref{eq:cov_v} are proportional to the eigenvalues $\lambda_i^c$ of the matrix:
\begin{equation}
    \tilde{\mathbf{\Delta}}= \left[\frac{d\mathbf{z}(t_f)}{d\mathbf{z}_0}\right]^T\left[\frac{d\mathbf{z}(t_f)}{d\mathbf{z}_0}\right]
\end{equation}
with $\mathbf{z}(t_f)\approx \sum_i^m \mathbf{c}_{i}\Psi_i(\mathbf{p})$, $\mathbf{p}=\mathbf{z}_0$, $m=1$ and $\Psi(\mathbf{z}_0)$ the Chebyshev polynomials of type 2.
\end{proposition}
\begin{proof}
Considering a first order expansion of $\mathbf{z}(t_f)\approx \sum_i^1 \mathbf{c}_{i}\Psi_i(\mathbf{z}_0)$. One can derive the matrix $\tilde{\Delta}$:
\begin{equation}
\mathbf{\tilde{\Delta}} =\left[\begin{array}{ccc}
\sum_{i=1} c_{1,i}^2(t)&...&\sum_{i=1} 
c_{1,i}(t)c_{n,i}(t)\\
...&...&...\\
...&\sum_{i=1} c_{j,i}^2(t)&...\\
...&...&...\\
\sum_{i=1} c_{n,i}(t)c_{1,i}(t)&...&\sum_{i=1} c_{n,i}^2(t)\\
\end{array}
\right]
\end{equation}
where the index $j$ loops over the number of dimensions $n$. From the definition of the covariance in \eqref{eq:cov_v} the terms in the summation are multiplied times the factors $s_i$ which descend from the integration over $\Omega$ of the product of basis functions. Assuming that the uncertainty in the components of $\mathbf{z}_0$ is independent and uncorrelated and that also in the covariance the polynomial expansion is up to first order, all terms $s_i$ have the same value $\tilde{s}$ and thus we can write:
\begin{equation}
    \mathbf{C}_v=\tilde{s}\tilde{\mathbf{\Delta}}
\end{equation}
\end{proof}
\begin{remark}
From linear algebra the Cauchy-Green deformation tensor $\mathbf{\Delta}= \left[\frac{d\mathbf{z}(t_f)}{d\mathbf{z}_0}\right]^T\left[\frac{d\mathbf{z}(t_f)}{d\mathbf{z}_0}\right]$ has the same eigenvalues of the matrix $\tilde{\mathbf{\Delta}}$, thus for a first order expansion with respect to $\mathbf{p}=\mathbf{z}_0$ it can be concluded that the eigenvalues used in the computation of the pseudo-diffusion exponent are proportional to the eigenvalues of the Cauchy-Green deformation tensor.   
\end{remark}
\begin{remark}
For infinitesimally small uncertainty in the initial conditions an expansion up to the first order is often a reasonable approximation and is consistent with a first order Taylor expansion of $\mathbf{z}(t_f)$ with respect to $\mathbf{z}_0$. If a higher order expansion is used instead the matrix $\tilde{\mathbf{\Delta}}$ will contain products of higher order coefficients and also the terms $s_i$ in the covariance will correspond to higher order polynomials. Thus an extension of Proposition \ref{pro:initial_cond_alpha} is not straightforward, however, one can notice that if the expansion is convergent the contribution of higher order terms will be small and the linear approximation in Proposition \ref{pro:initial_cond_alpha} will capture the main contribution to the value of the eigenvalues.
\end{remark}
Note that although in this paper we limited our attention only to the case of parametric uncertainty the same methodology can be applied to the study of dynamical systems driven by stochastic processes via the Karhunen-Lo\`eve expansion \cite{deheuvels}.
\subsection{Relation to Other Indicators Derived from Polynomial Expansions}
In \cite{Palau2015}, two dynamical indicators were derived from Jet Transport. One indicator was measuring the rate of contraction or expansion of the region propagated with Jet Transport. The rate was calculated with respect to the size of the set of initial conditions that was propagated.
In the definition of $\tilde{\alpha}$, as demonstrated in Proposition \ref{pro:initial_cond_alpha}, the set of initial conditions is $\Omega$ and a measure of its size is accounted for in the integrals of the polynomial bases, see  \eqref{eq:quadrature_norm}. The expansion/contraction is directly measured by the eigenvalues of the covariance matrix $\mathbf{C}_v$. In fact, given a covariance matrix, the ellipsoid enclosing a given percentile of the propagated states has the direction of its axes defined by the eigenvectors of $\mathbf{C}_v$ and their lengths by $2c\sqrt{\lambda_i}$, where $\lambda_i$ are the eigenvalues and $c$ defines the percentile. In Proposition \ref{pro:initial_cond_alpha} we also demonstrated that the eigenvalues of $\mathbf{C}_v$ are scaled by the integral of the basis functions over $\Omega$. Thus it can be concluded that if the pseudo-diffusion exponent is used to quantify the uncertainty in the propagated states from a set of uncertain initial conditions, it contains the same information, on the expansion or contraction of the initial uncertainty set, as the contraction/expansion indicator proposed in \cite{Palau2015}. 

In \cite{Iosto2022} an indicator was derived from the magnitude of the predicted and observed coefficients of a polynomial expansion of the propagated states. This indicator was called "n+1". As it was argued above, SFTLE2 is, by its nature, capturing the variation in the high order coefficients of the polynomial expansion and is, therefore, related to the n+1 indicator. In fact it was shown that irregular types of motion require higher order expansions to achieve a good accuracy of the polynomial representation. At the same time neighbouring initial conditions are shown to lead to different evolutions of the polynomial expansions when two trajectories tend to diverge. In this sense the SFTLE2 is also connected to the indicator, proposed in \cite{Palau2015}, measuring the precision of the polynomial expansion of the propagated states. However, SFTLE2 presents two important differences:i) SFTLE2 is not suitable to quantify the uncertainty in the initial conditions because the difference of the coefficients is computed with respect to an infinitesimal variation of the initial conditions; ii) SFTLE2 encapsulates both a measure of the divergence of two neighbouring trajectories and a measure of the uncertainty in the propagated states induced by model uncertainty.

\section{Practical Utility of the Indicators}
In this section we present two practical uses of the proposed indicators. The first practical use is the identification of robust initial conditions in the Elliptical Restricted Three-Body Problem. We will give a definition of robust initial conditions and show how $\tilde{\alpha}$ can be used to design trajectories that are weakly affected by the uncertainty in the dynamic model. The second practical use is the identification of regions of practical stability in the CR3BP.
For all calculations in this section polynomials were expanded to order 3 and 9 abscissa points per dimension of the uncertain vector $\mathbf{p}$ were used. The expectation $\mathbb{E}$ was computed by drawing 100 uniformly distributed samples from the space $\Omega$ and evaluating the polynomial chaos at $t_f$.

\subsection{Identification of Robust Initial Conditions}
\begin{figure}[htb]
     \centering
     \subfloat[$\tilde{\alpha}$]{\includegraphics[width=0.5\textwidth]{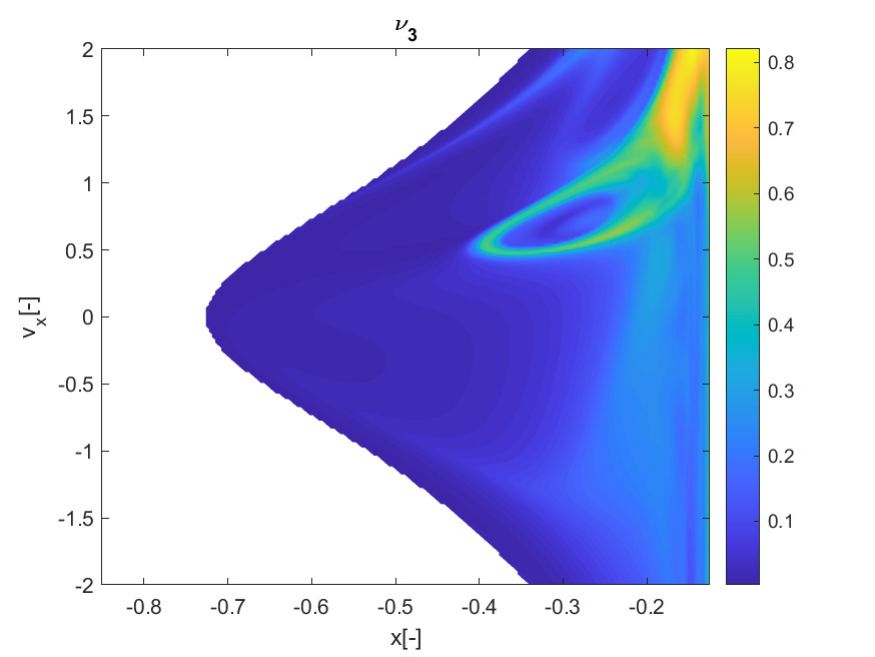}}
     \subfloat[$\mathbb{E}_{0.1}$]{\includegraphics[width=0.5\textwidth]{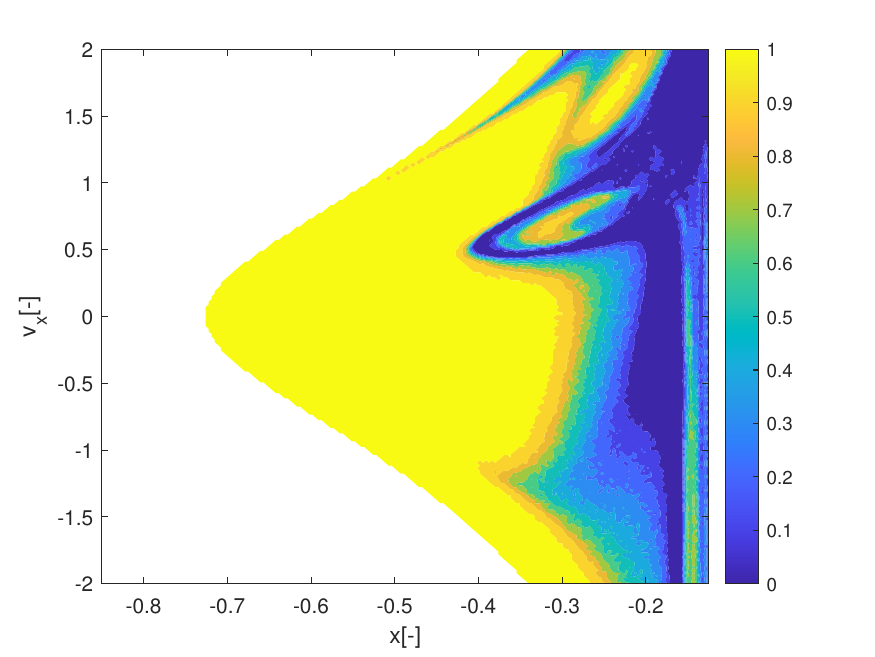}}\\
     \caption{Pseudo-diffusion exponent for the ER3BP.}
     \label{fig:e3bp_alpha}
\end{figure}

As previously mentioned, the major utility of the indicators proposed in this paper is to study the effect of model uncertainty on the evolution of a trajectory starting from a given initial condition $\mathbf{z}_0$. 
For example, in \cite{gaw2007}, the authors studied how Lagrange Coherent Structures would change due to a variation of the eccentricity in the Elliptical Restricted Three Body Problem. We can understand this variability as an uncertainty in the existence and location of the LCS induced by an uncertainty in the eccentricity. The whole study in \cite{gaw2007} can be revisited by computing the SFTLE1, which would quantify the effect of the uncertainty in $e$ on the FTLE. A low value of SFTLE1 would correspond to initial conditions that display a low sensitivity to a variation of the eccentricity.
The same logic can be applied to the pseudo-diffusion exponent as, for a given initial condition, $\tilde{\alpha}$ would be small if the trajectories in an ensemble presented a small variance with respect to a variation of the eccentricity. 
Following this idea, we define the robustness of a given initial condition $\mathbf{z}_0$ as:
\begin{definition}
\label{def:robustness}
The initial condition $\mathbf{z}_0$ is robust, with respect to the uncertainty vector $\mathbf{p}$, with robustness index $\rho_p$, if $\bar{\alpha}<\rho_p$, where $\bar{\alpha}=\tilde{\alpha}$, if the pseudo-diffusion exponent is used to study the dynamics, or $\bar{\alpha}=\alpha_1^2$, if SFTLE1 is used to study the dynamics instead. Therefore, minimum $\rho_p$ initial conditions maximise robustness with respect to the uncertainty in $\mathbf{p}$.
\end{definition}
Consider now the case in which a mission analyst needs to identify minimum control trajectories in a binary system with poorly known physical parameters. This is, for example, the case of missions to binary asteroids. Given the limited knowledge of the exact mass of the asteroids and the uncertainty in the orbital parameters of the secondary, there is an interest in finding initial conditions that are robust with respect to the uncertainty in the physical parameters of the binary system.
Definition \ref{def:robustness} can be straightaway applied to this case. As an illustrative example, consider the problem of finding robust initial conditions in the uncertain Elliptical Restricted 3-Body Problem (ER3BP). Following \cite{gaw2007} and \cite{Palau2015} the planar ER3BP Problem can be modelled as follows:
\begin{equation}
    \label{eq:e3bp1ode}
        \mathbf{z}' = \frac{d \mathbf{z}}{d \theta_s}
    \begin{array}{c}
\end{array}
=
    \begin{array}{c}
\begin{bmatrix}
v_x \\
v_y \\
2 v_y + \frac{\partial J}{\partial x}/(1+e\cos\theta_s) \\
- 2 v_x + \frac{\partial J}{\partial y}/(1+e\cos\theta_s)
\end{bmatrix}
\end{array}
\end{equation}
where $e$ is the eccentricity of the orbit of the secondary body, $\mathbf{z}=[x,y,v_x,v_y]^T$ and $\theta_s$ its true anomaly.
As in \cite{Palau2015}, we use the pseudo-energy:
\begin{equation}
\label{eq:energy_e}
    E(x, y, v_x, v_y) = \frac{1}{2} (v_x^2 + v_y^2) - J(x, y)/(1+e\cos\theta_s) \end{equation}
    
to reduce the number of free initial conditions and define the value of $v_y$ as:

\begin{equation}
    \label{eq:ydot_e}
    v_y = - \sqrt{2(E_0 + J/(1+e\cos\theta_{s0})) - v_x^2}
\end{equation}
with $\theta_{s0}=0$. We then consider an uncertainty in both the eccentricity $e$ and the mass parameter $\mu$ (see Table \ref{tab:e3bp_param}) around the values in the examples presented in \cite{Palau2015}.
Figs.\ref{fig:e3bp_alpha} show the pseudo-diffusion exponent and the expectation for $\epsilon=0.1$. The $\tilde{\alpha}$ looks similar to the one of the CR3BP, however, Fig.\ref{fig:e3bp_alpha}b displays an interesting area in the upper right corner that is less pronounced in the case of the CR3BP. 

\begin{figure}[htb]
     \centering
     \subfloat[$\tilde{\alpha}<0.01$]{{\includegraphics[width=0.5\textwidth]{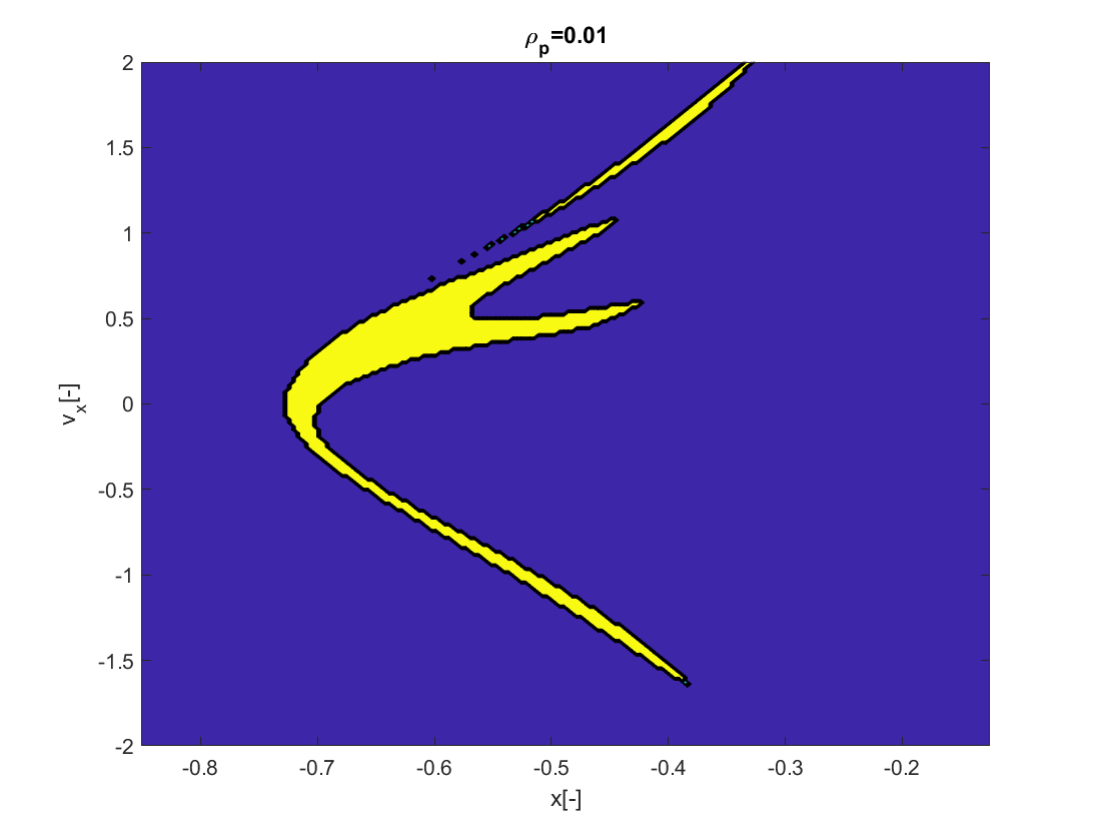}}}
     \subfloat[$0.4<\tilde{\alpha}<0.6$]{{\includegraphics[width=0.5\textwidth]{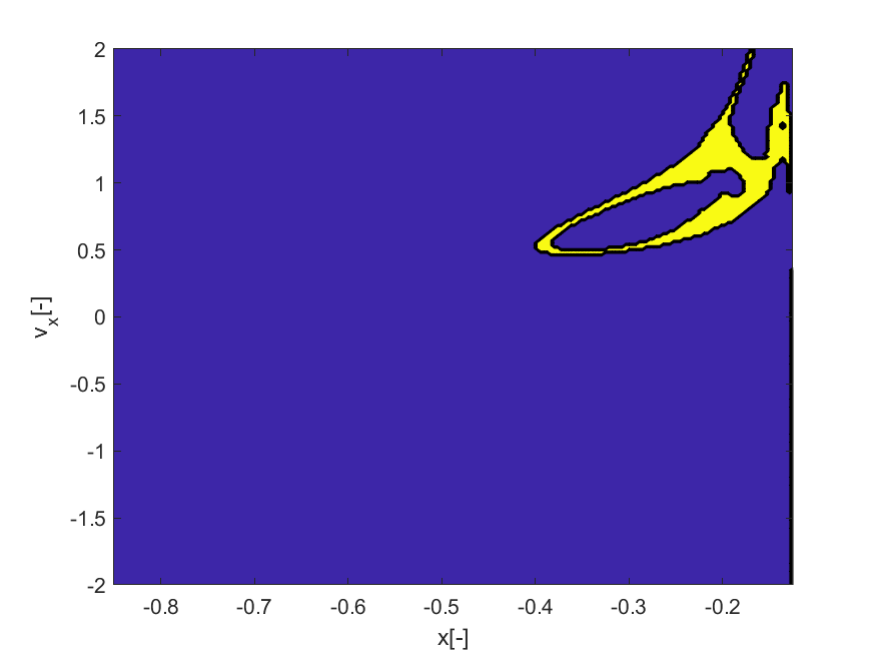}}}\\
     \caption{Robustness regions for the ER3BP.}
     \label{fig:e3bpxy_rhop}
\end{figure}

\begin{figure}[htb]
     \centering
     \subfloat[]{{\includegraphics[width=0.55\textwidth]{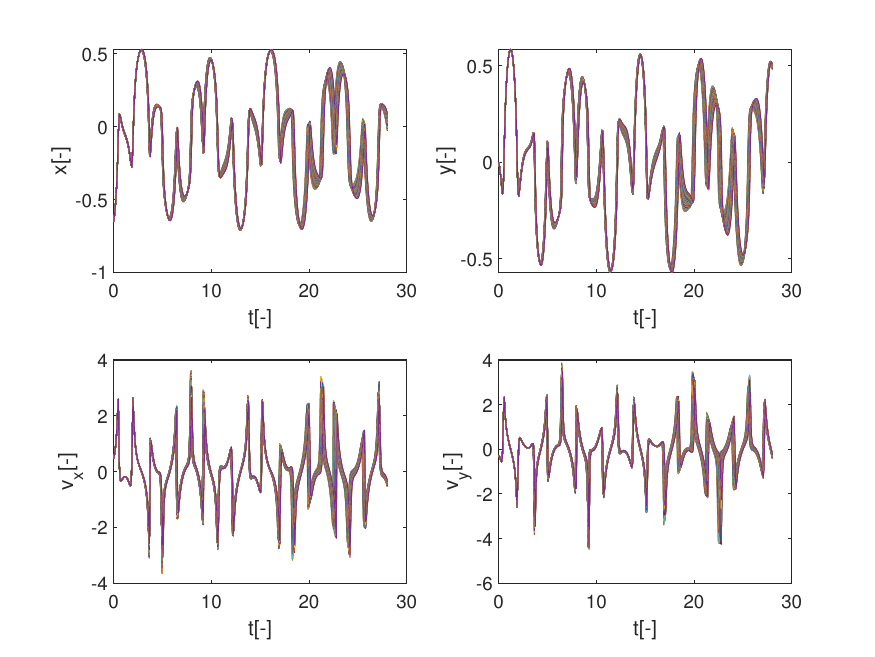}}}
     \subfloat[]{{\includegraphics[width=0.55\textwidth]{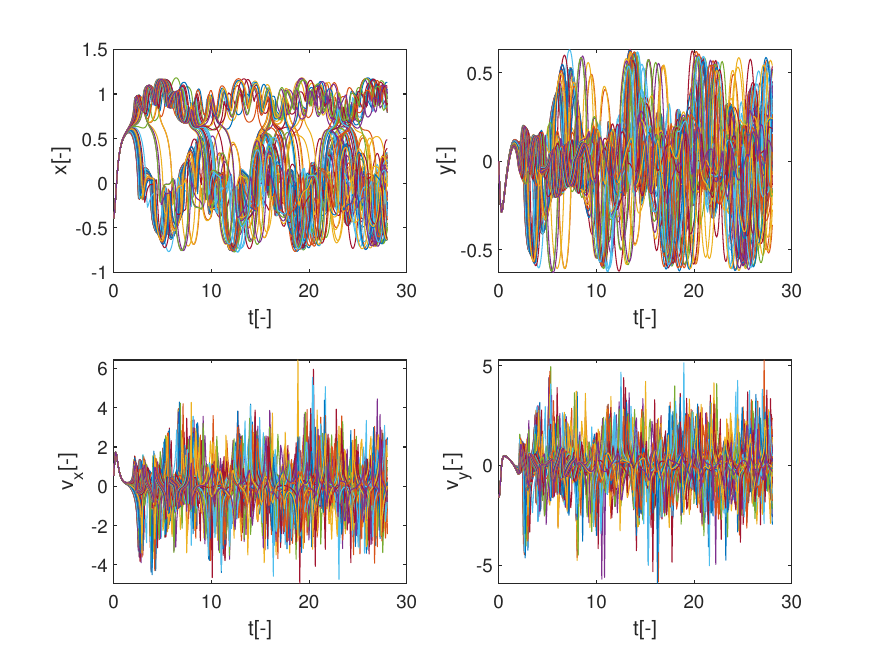}}}\\     
     \caption{Example of ensemble of trajectories for: a) $\tilde{\alpha}<0.01$ and, b) $0.4<\tilde{\alpha}<0.6$. Integration time $t_f=28$.}
     \label{fig:e3bp_components_alpha}
\end{figure}

\begin{table}[]
    \centering
        \caption{Summary of parameter settings for the ER3BP}
    \begin{tabular}{|c|c|c|}
         \hline
         Eccentricity & Mass & Integration\\
         & Parameter & Time\\
         \hline
         $e\in[0.04-10^{-3},0.04+10^{-3}]$ & $\mu\in [0.1-10^{-3},1+10^{-3}]$& $t_f=2.8$\\
         \hline
    \end{tabular}
    \label{tab:e3bp_param}
\end{table}
Figs. \ref{fig:e3bpxy_rhop} show the initial conditions for which $\tilde{\alpha}$ is respectively below 0.01 and within the interval $[0.4,0.6]$ for the ER3BP. 
Figs.\ref{fig:e3bp_components_alpha}a and b show two ensembles of 81 trajectories, one starting respectively from initial condition $\mathbf{z}_0=[-0.416457, 1.51759]$ belonging to the region identified in Fig. \ref{fig:e3bpxy_rhop}a  and the other starting from initial condition $\mathbf{z}_0=[-0.390955, 0.532663]$ belonging to the region identified in Fig. \ref{fig:e3bpxy_rhop}b. From Figs. \ref{fig:e3bp_components_alpha} one can see how the ridges identified by the pseudo-diffusion exponent correspond to ensembles of trajectories that start from the same identical initial condition but due to the effect of uncertainty display very different behaviours and diverge quite quickly. On the contrary regions of low $\tilde{\alpha}$ corresponds to ensembles where trajectories remain close to each other.
\subsection{Identification of Practical Stability Regions of the CR3BP}
\begin{figure}
     \centering
     \subfloat[]{{\includegraphics[width=0.5\textwidth]{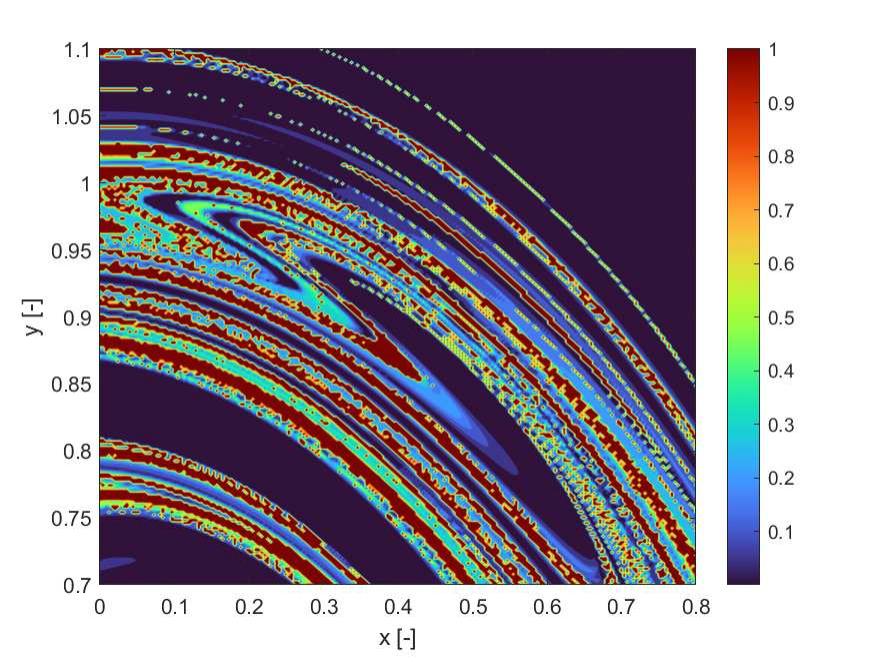}}}
     \subfloat[]{{\includegraphics[width=0.5\textwidth]{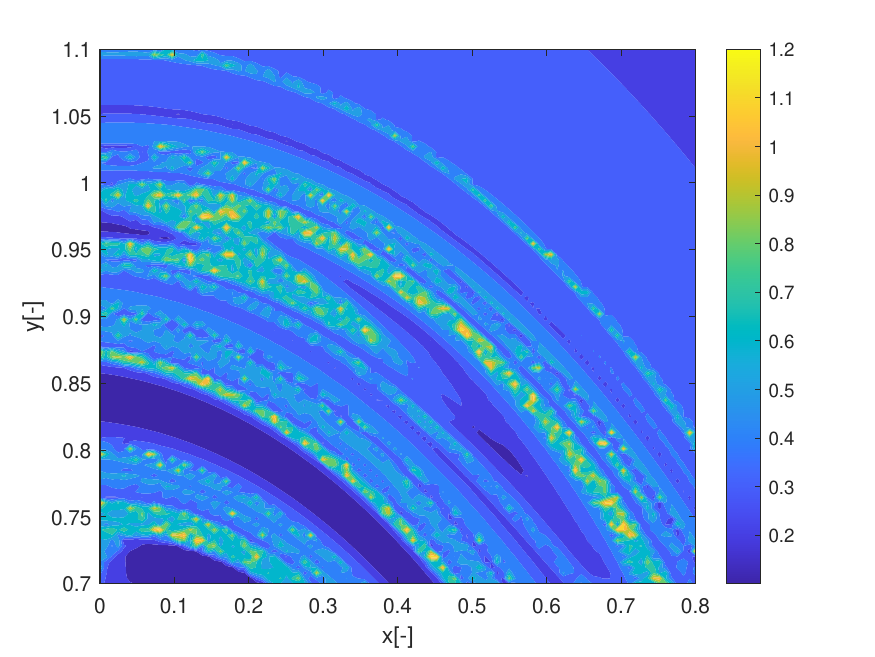}}}\\
     \subfloat[]{{\includegraphics[width=0.5\textwidth]{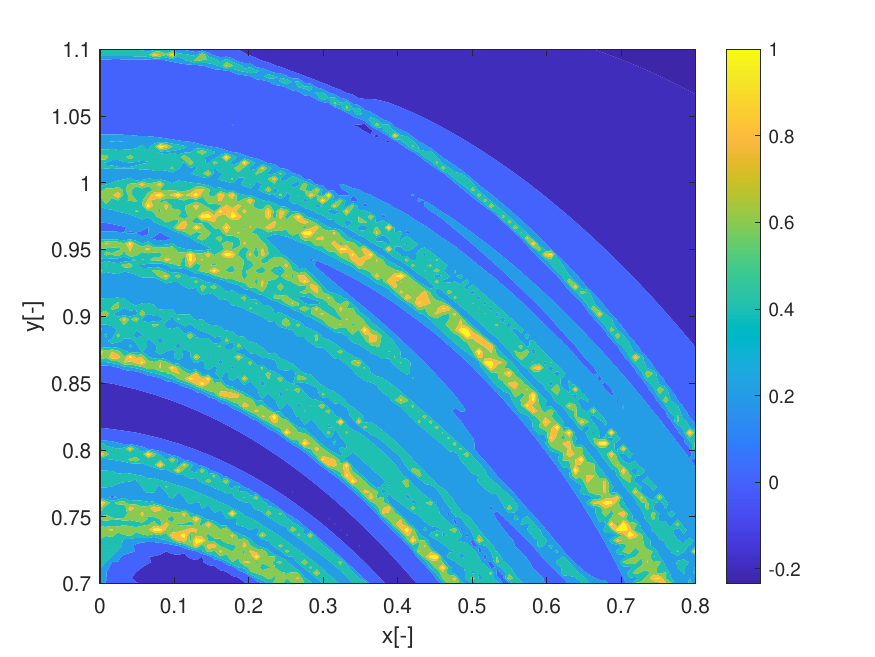}}}
     \subfloat[]{{\includegraphics[width=0.5\textwidth]{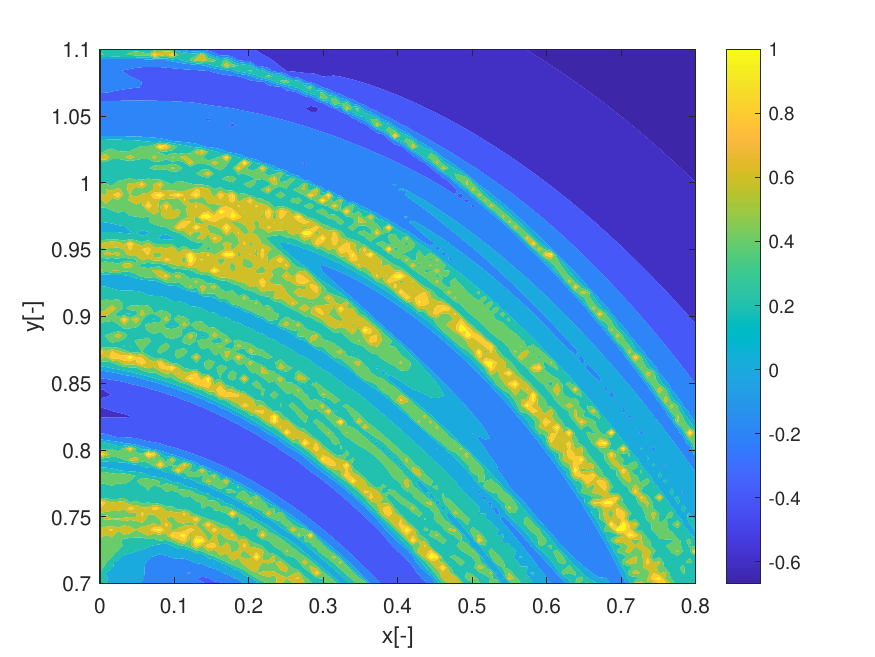}}}\\
     \subfloat[]{{\includegraphics[width=0.5\textwidth]{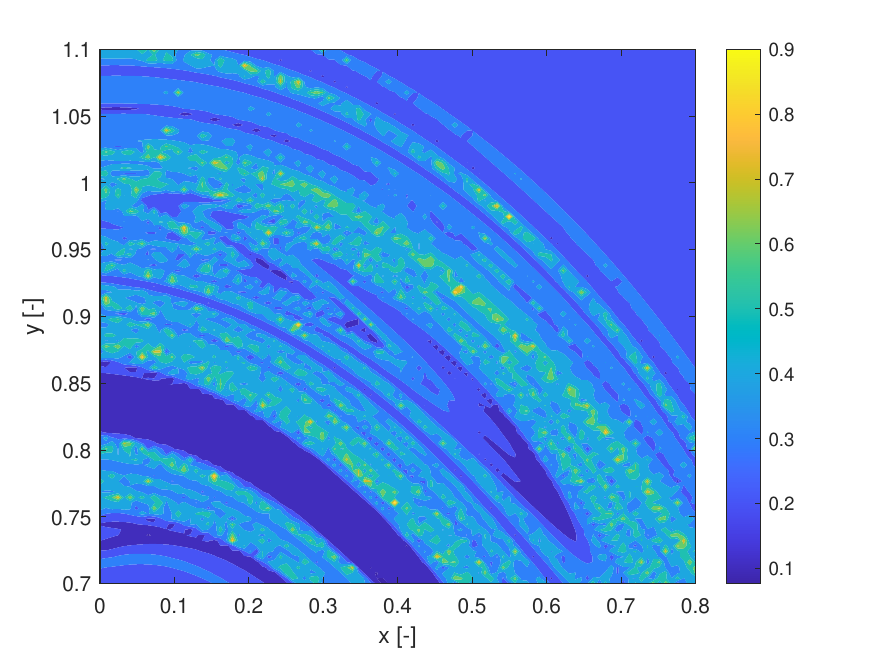}}}     
     \subfloat[]{{\includegraphics[width=0.49\textwidth]{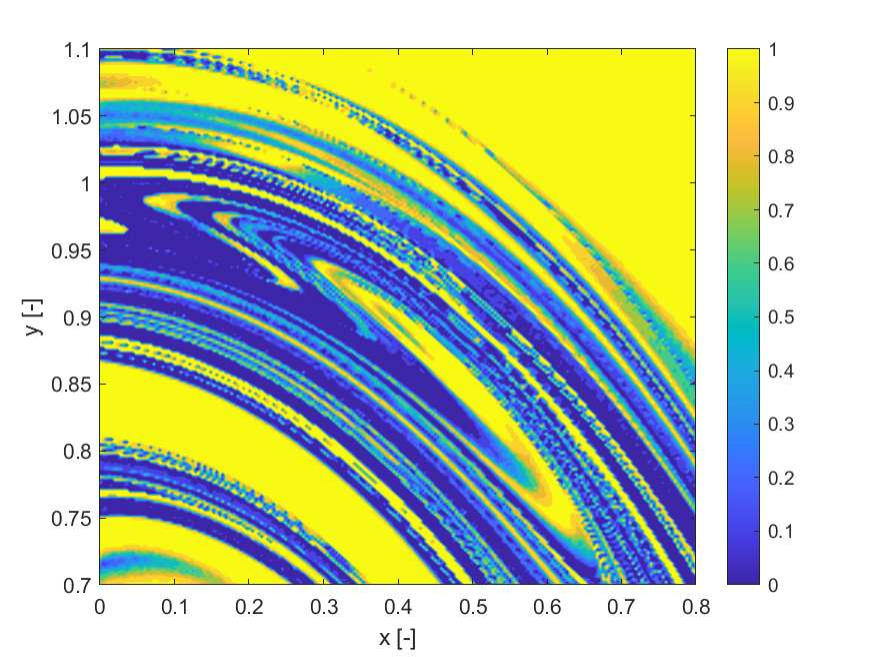}}}\\     
     \caption{Stability regions around L4 in the CR3BP: a) $\tilde{\alpha}$ and, b) $\sigma_2^1$, c) $\sigma_2^2$, d) $\sigma_2^3$.e) FTLE, f) $\mathbb{E}_{0.1}$ Integration time $t_f=20$.}
     \label{fig:3bp_stability}
\end{figure}

\begin{figure}
     \centering
     \subfloat[]{{\includegraphics[width=0.5\textwidth]{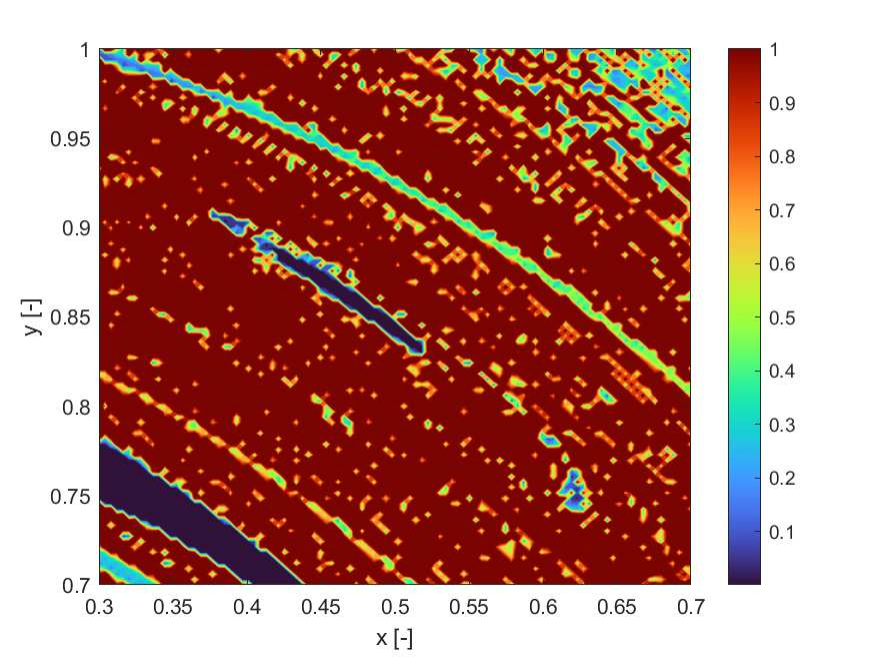}}}
     \subfloat[]{{\includegraphics[width=0.5\textwidth]{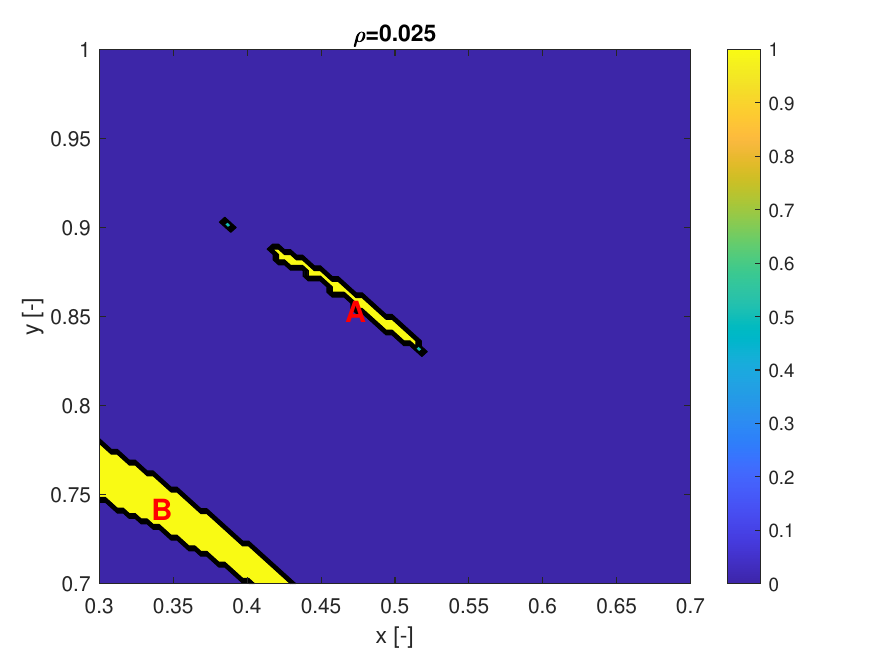}}}\\     
     \subfloat[]{{\includegraphics[width=0.5\textwidth]{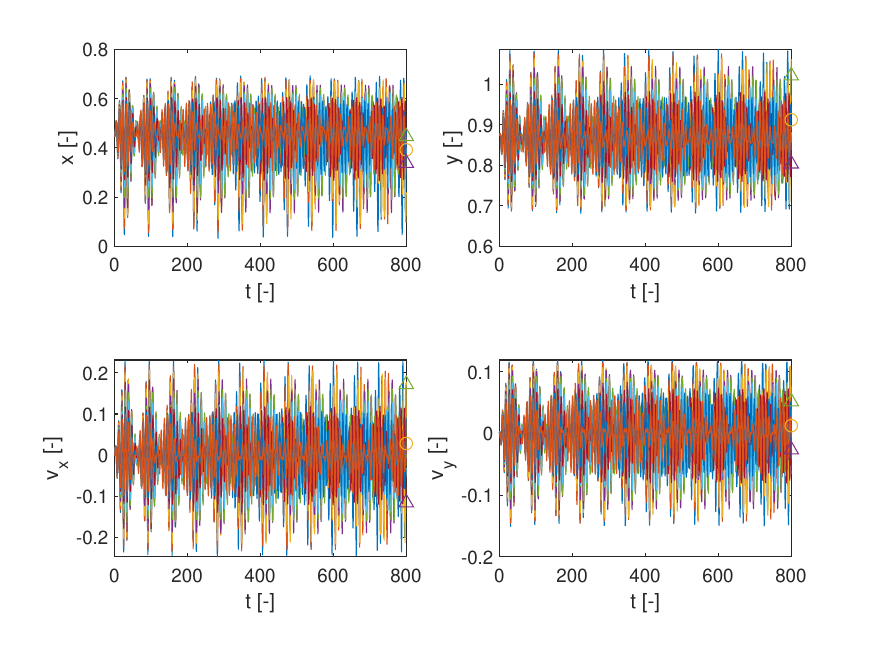}}}
     \subfloat[]{{\includegraphics[width=0.5\textwidth]{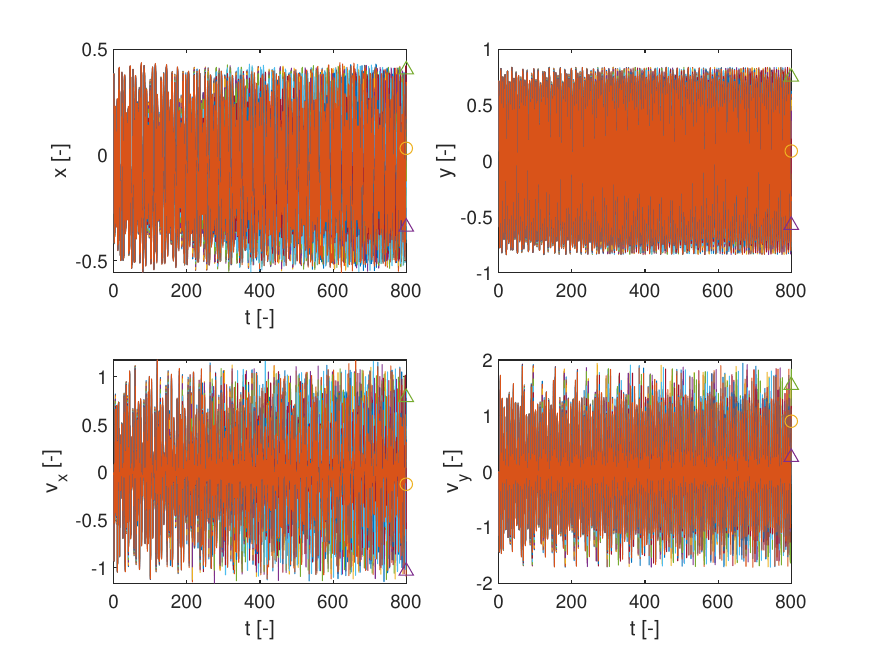}}}\\
     \subfloat[]{{\includegraphics[width=0.5\textwidth]{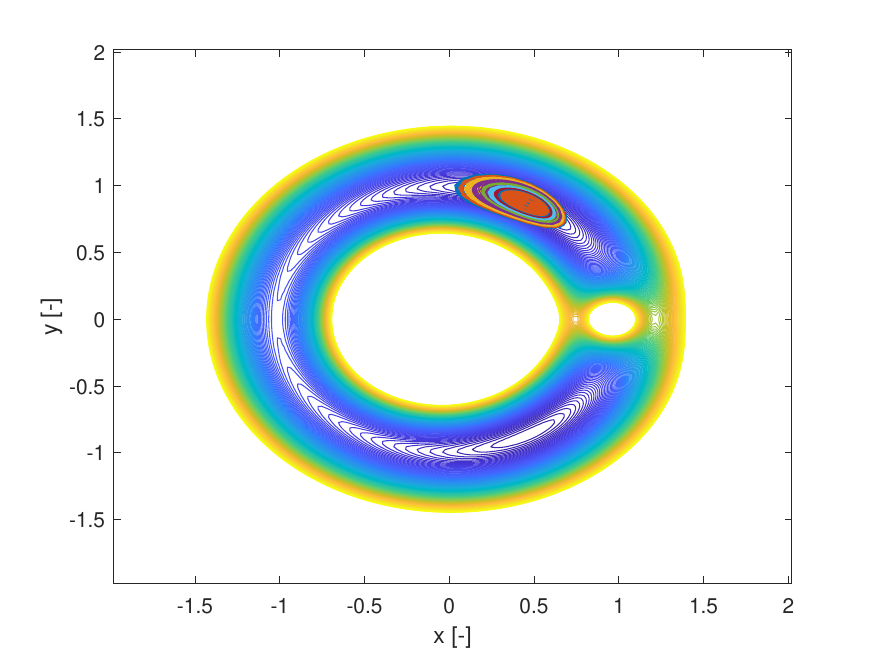}}}
     \subfloat[]{{\includegraphics[width=0.5\textwidth]{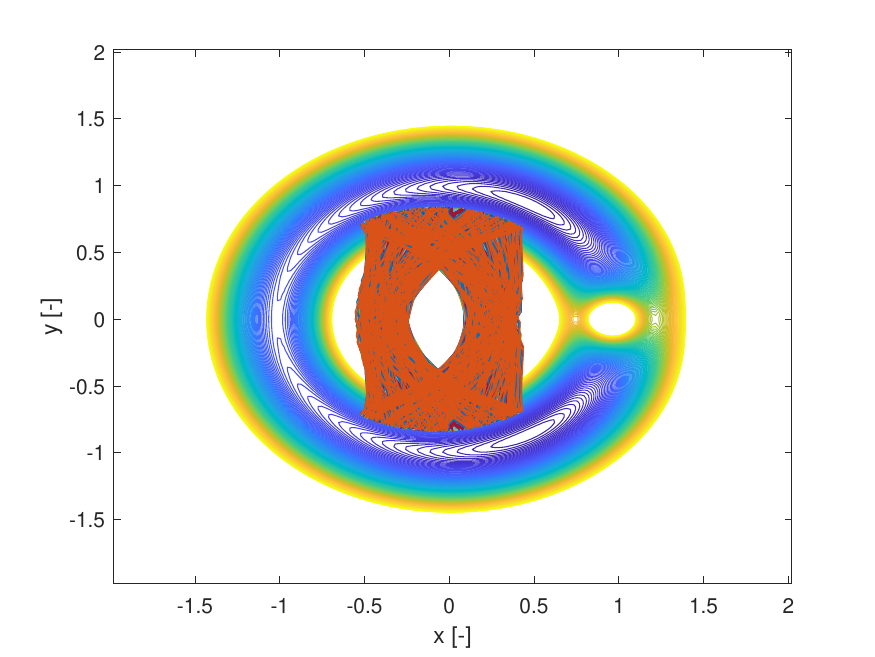}}}\\
     \caption{Close up of stability regions around L4 in the CR3BP for extended integration time $t_f=80$: a) $\tilde{\alpha}$ and, b) $\tilde{\alpha}<0.025$, c) components for sample taken from region A for integration time $t_f=800$, d) components for sample taken from region B for integration time $t_f=800$, e) trajectory of sample taken from region A for integration time $t_f=800$, f) trajectory of sample taken from region B for integration time $t_f=800$.}
     \label{fig:3bp_stability_closeup}
\end{figure}

In this section we show how the indicators proposed in this paper can be used to identify regions of practical stability in the CR3BP in the case in which the model of the dynamical system is uncertain. The analyses in this section extends the one in \cite{Palau2015} in that the dynamics is considered uncertain and thus it reflects more closely the situation in which a space mission to a new binary system is designed. The indicator are calculated for different values of the initial condition $\mathbf{z}_0=[x_0,y_0,0,0]^T$ assuming an uncertainty in the mass parameter. The expected value of the mass parameter is chosen to be $\mu=0.039$ which is slightly above the limit of the linear stability condition for the triangular points. We then considered an uncertainty on the value of the mass parameters so that $\mu\in[0.039-10^{-3},0.039+10^{-3}]$. Thus for some realisations of $\mu$ the triangular points are linearly stable and for others are not. The question is whether there are regions around L5 and L4 that provide practical stability for all realisations of the uncertain parameter.
Figs. \ref{fig:3bp_stability} show the regions around L4 identified by the pseudo-diffusion exponent and the SFTLE2 of the first three coefficients of the polynomial expansion. In Fig. \ref{fig:3bp_stability}a one can read the value of $\tilde{\alpha}$ for an integration time $t_f=20$. Dark blue means low diffusion, while all values equal to 1 (red regions) imply that at least one realisation has a collision with one of the two primary bodies. Figs. \ref{fig:3bp_stability}b,c and d show, respectively, $\sigma_2^1$,$\sigma_2^2$ and $\sigma_2^3$ while Fig. \ref{fig:3bp_stability}e shows the FTLE for the nominal value $\mu=0.039$.
Finally Fig. \ref{fig:3bp_stability}f shows the expectation for $\epsilon=0.1$. In this last case yellow regions correspond to low diffusion and no collisions.
At this point one might want to know if the solutions that appear to be practically stable for $t_f=20$ can be extended for longer integration times. To this end we analysed a smaller region around L4. We restricted the range of values of x and y to the intervals $[0.3,0.7;0.7,1.0]$, extended the integration time to $t_f=80$ and re-calculated $\tilde{\alpha}$. The result can be seen in Figs.\ref{fig:3bp_stability_closeup}. Fig. \ref{fig:3bp_stability_closeup}a is the value of the pseudo-diffusion exponent where values of 1 correspond to collisions of at least one trajectory in the ensemble. In Fig. \ref{fig:3bp_stability_closeup}b we isolated only the regions for which $\tilde{\alpha}<0.025$. We then took two random initial conditions from region A and region B in Fig. \ref{fig:3bp_stability_closeup}b and integrated, from those initial conditions, an ensemble of trajectories, for $t_f=800$. In particular the two samples have initial conditions $x=0.446231$, $y=0.874874$, in region A, and $x=0.384848$, $y=0.718182$, in region B. The individual components and the corresponding trajectory ensemble in configuration space are represented in Figs. \ref{fig:3bp_stability_closeup}c and e, and d and f respectively.
Note that region B was identified also in \cite{Palau2015} that also presents an example of trajectory similar to the one in \ref{fig:3bp_stability_closeup}f. 


\section{Conclusions \& Future Work}
\label{sec:concl}
This paper introduced three indicators that quantify the effect of parametric uncertainty on the time evolution of nonlinear dynamical systems. Two are derived from the concept of Finite Time Lyapunov Exponents and one from the relationship between mean square displacement and time in the case of anomalous diffusion. It was shown how the three indicators provide consistent information on the dynamics when used to build a cartography of the phase space. 

While SFTLE1 simply quantifies the statistical moments of the standard FTLE, the other two indicators were shown to relate the time evolution of the coefficients of polynomial expansions with the chaotic and diffusive nature of the motion. It was also experimentally and theoretically demonstrated that the quantification of the uncertainty in the initial conditions is equivalent to the computation of the FTLE when this uncertainty is infinitesimal.

The paper presented a measure of the probability associated to the diffusion of an ensemble of trajectories. At the same time it was argued that the weight function does not need to be a probability distribution. Any orthogonal polynomials with respect to any weight function can be used. More in general any form of polynomial-based quantification of uncertainty, whether intrusive or non-intrusive, can be used provided that the polynomials can be orthogonalised. 

The computational complexity of the calculation of these indicators is mainly related to the complexity of the propagation of uncertainty with polynomials. On the other hand it was shown that the pseudo-diffusion exponent has lower computational complexity for the same number of uncertainty parameters because it does not require the propagation of the variational equations. Note that in this paper the indicators were computed for a particular final times $t_f$ but could be equally computed for multiple times $t$ to study their evolution.

From a practical applicability standpoint, it was shown how the indicators could be used to find sets of robust initial conditions and the pseudo-diffusion indicator could identify regions of practically stable trajectories around L4, in the CR3BP, also in the case in which the uncertainty in the mass parameter would imply that the triangular points are linearly unstable. 

Future work will further extend these indicators to account for stochastic processes driving the dynamical systems and imprecision in the distribution functions.

\section*{Acknowledgements}
This work was supported through the MSCA ETN Stardust-R grant agreement number 813644.


\bibliography{Biblio.bib}   

\end{document}